\documentclass{emulateapj}
\usepackage{apjfonts}
\usepackage{lscape}
\begin{document}
\shorttitle{SAGE Survey LMC PNe}
\shortauthors{Hora et al.}
\slugcomment{To be published in the Astronomical Journal; accepted 2007 October 30}

\title{SPITZER SAGE Observations of Large Magellanic Cloud Planetary
Nebulae} 

\author{ J. L. Hora\altaffilmark{1},
M. Cohen\altaffilmark{2},
R. G. Ellis\altaffilmark{3},
M. Meixner\altaffilmark{4},
R. D. Blum\altaffilmark{5},
W. B. Latter\altaffilmark{6},
B. A. Whitney\altaffilmark{7},
M. R. Meade\altaffilmark{8},
B. L. Babler\altaffilmark{8},
R. Indebetouw\altaffilmark{9},
K. Gordon\altaffilmark{10},
C. W. Engelbracht\altaffilmark{10},
B.--Q. For\altaffilmark{10}, 
M. Block\altaffilmark{10}, 
K. Misselt\altaffilmark{10}, 
U. Vijh\altaffilmark{4}, 
C. Leitherer\altaffilmark{4}}

\altaffiltext{1}{Center for Astrophysics, 60 Garden St., MS 65, 
Cambridge, MA 02138} 
\altaffiltext{2}{Radio Astronomy Laboratory, 601 Campbell Hall, University
of California at Berkeley, Berkeley, CA 94720} 
\altaffiltext{3}{Brown University, Providence, RI 02912}
\altaffiltext{4}{Space Telescope Science Institute, 3700 San Martin
Way, Baltimore, MD 21218} 
\altaffiltext{5}{National Optical Astronomy Observatory, 950 North 
Cherry Ave., Tucson, AZ, 85719} 
\altaffiltext{6}{Caltech, NASA Herschel Science Center, MS
100--22, Pasadena, CA 91125} 
\altaffiltext{7}{Space Science Institute, 4750 Walnut St. Suite 205, 
Boulder, CO 80301, bwhitney@spacescience.org}
\altaffiltext{8}{Department of Astronomy, 475 North Charter St., 
University of Wisconsin, Madison, WI 53706}
\altaffiltext{9}{Dept. of Astronomy, University of Virginia, P.O. Box 
3818, Charlottesville, VA 22903}
\altaffiltext{10}{Steward Observatory, University of Arizona, 933 North
Cherry Ave., Tucson, AZ 85719}

\begin{abstract}

We present IRAC and MIPS images and photometry of a sample of previously
known planetary nebulae (PNe) from the SAGE survey of the Large Magellanic
Cloud (LMC) performed with the Spitzer Space Telescope.  Of the 233 known
PNe in the survey field, 185 objects were detected in at least two of the
IRAC bands, and 161 detected in the MIPS 24 $\mu$m images.  Color-color
and color-magnitude diagrams are presented using several combinations of
IRAC, MIPS, and 2MASS magnitudes.  The location of an individual PN in the
color-color diagrams is seen to depend on the relative contributions of
the spectral components which include molecular hydrogen, polycyclic
aromatic hydrocarbons (PAHs), infrared forbidden line emission from the
ionized gas, warm dust continuum, and emission directly from the central
star. The sample of LMC PNe is compared to a number of Galactic PNe and
found to not significantly differ in their position in color-color space.
We also explore the potential value of IR PNe luminosity functions (LFs)
in the LMC. IRAC LFs appear to follow the same functional form as the
well-established [\ion{O}{3}] LFs although there are several PNe with
observed IR magnitudes brighter than the cut-offs in these LFs.  
\end{abstract}

\keywords{planetary nebulae: general --- Magellanic Clouds --- infrared: 
stars --- stars: mass loss}

\section{Introduction} 

The Large Magellanic Cloud (LMC) has been important for the study of many
astrophysical processes and objects because it is one of the nearest
galaxies to our own, and due to its location above the Galactic plane and
its favorable viewing angle \citep[35$\degr$;][]{vanderm01}, the system
can be relatively easily surveyed and many of its global properties
determined.  These properties are important in particular for the study of
planetary nebulae (PNe).  The known distance to the LMC removes the
relatively large uncertainty in this parameter that affects many Galactic
PNe \citep{hajian06}.  The distance of $\sim$ 50 kpc allows individual
objects to be isolated and in some cases resolved.  The effects on PNe of
the lower metallicity and dust/gas mass ratio in the LMC can be explored.  
One can also hope to detect a large fraction of the total number of PNe in
the LMC, as opposed to in the Galaxy, where confusion and extinction in
the plane allow us to detect only about 10\% of the PNe expected to exist
\citep{kwok00,frew05}.

An infrared survey of the LMC called Surveying the Agents of a Galaxy's
Evolution \citep[][SAGE]{meixner06} has recently been completed using the
IRAC \citep{fazio04} and MIPS \citep{rieke04} instruments on the Spitzer
Space Telescope \citep{werner04}.  SAGE is an unbiased, magnitude-limited
survey of a $\sim 7\degr \times 7\degr$ region centered on the LMC.  This
Spitzer ``Legacy'' survey has provided a tremendous resource for the study
of the stellar populations and interstellar medium (ISM) in the LMC.   
Some early results on the evolved stellar populations were given by
\citet{blum06}, who identified $\sim$32,000 evolved stars brighter than
the red giant tip, including oxygen-rich, carbon-rich, and ``extreme''
asymptotic giant branch (AGB) stars.

In this paper we explore the properties of a sample of known PNe as
revealed by the SAGE data.  The catalog of 277 LMC PNe assembled by
\citet{leisy97} from surveys that cover an area of over 
100 square degrees was 
used for the source of positions of the PNe.  Leisy et al.
used CCD images and scanned optical plates to obtain accurate positions of
the objects to better than 0\farcs5.  They point out that the
objects are in general PN candidates, with only 139 confirmed at that time
with slit spectroscopy.  For simplicity we will refer to the objects in the 
catalog as PNe, even though this caveat still applies for many of the sources.
When we began to work with the SAGE data, the
Leisy et al. catalog was the largest summary list of the known PNe at the
time.  During the course of this work, \citet{reid06} published a list of
PNe in the central 25 deg$^2$ of the LMC, including 169 of the previously
known objects and 460 new possible, likely, or true PNe. We will present
our results here for the \citet{leisy97} catalog, and a future paper will
include the new objects in the \citet{reid06} survey.

\section{Observations and Reduction}

The observations were obtained as part of the SAGE survey of the LMC
\citep{meixner06}. For the IRAC data, we did not use the SAGE catalog
directly since the catalog is constructed to contain point sources and
some of the PNe are likely to be extended in the IRAC images.  Also, when
we began this work both epochs had been taken but only the epoch 1 catalog
was available, so by making our own mosaics using both epochs and
performing the photometry we could obtain higher sensitivity and be less
susceptible to instrument artifacts and cosmic rays. Using the known
LMC PNe locations from the \citet{leisy97} catalog, all Basic Calibrated
Data (BCD) images within 6\arcmin\ of the known positions were collected
for inclusion in the mosaics. The SAGE survey area covered 233 of the
Leisy et al. LMC PNe positions.

\subsection{IRAC data}

We used the version 13.2.0 BCD images as the starting point in our reduction. 
The 13.2.0 version of
the pipeline had improved pointing reconstruction compared to previous
versions, but the ``DARKDRIFT'' module which normalizes possible detector
output channel offsets was turned off for the 3.6, 4.5, and 8.0 $\mu$m
channels (it has since been turned on for BCD versions 14 and beyond).  
Before further processing, the ``jailbar'' correction
algorithm\footnote{Available on the Spitzer Science Center contributed
software web pages at http://ssc.spitzer.caltech.edu/irac/jailbar} was
applied to the BCD to remove the ``pinstriping'' artifact possible in this
version of the data.  After applying this correction, the BCD images are
essentially the same as the S14 pipeline version.  The images were then
cleaned using custom IRAF\footnote{IRAF is distributed by the National
Optical Astronomy Observatories, which are operated by the Association of
Universities for Research in Astronomy, Inc., under cooperative agreement
with the National Science Foundation.} cleaning scripts to remove residual
striping, banding and column pulldown artifacts. The mosaicing process
removed any transient events such as cosmic rays and bad pixels as well as
minimizing any fixed-pattern background noise. The BCD were combined into
mosaics using the IRACproc post-BCD Processing package version 4.0
\citep{schuster06}. This package is based on the mopex mosaicing software
released by the Spitzer Science Center (SSC) \citep{makovoz06} but uses an
improved outlier detection method appropriate for low coverage that
rejects cosmic rays and other transients without removing pixels in the
cores of real point sources. A pixel size of 0\farcs6 and corresponding
subpixel alignment of the BCD was used for the individual images to
improve the resolution of the final mosaic and allow for the detection of
finer structures and separation of point sources.

The IRAC photometry was extracted from these mosaics using the IRAF
routines daofind and phot.  The closest matching IRAC source to the
\citet{leisy97} catalog position if less than 2 arcsec away
was assumed to be the PN.  In the 233
fields covered by the IRAC images, 185 PNe were detected in at least two
of the IRAC bands. There were 119 sources detected in all four bands, 19 in 
bands 1 and 2 only, one in bands 3 and 4 only, and 24 sources detected in 
only one band.   Since some of the fields were crowded with many point
sources, a relatively small aperture size (diameter of 2\farcs8) was used.  
The source crowding was more likely to affect the 3.6 and 4.5 $\mu$m images
since there is significant stellar continuum emission at those wavelengths from main sequence stars.  The 5.8 and 8.0 $\mu$m images were more likely to be
affected by extended structured emission from the  ISM in the LMC, especially
in areas near star-forming regions.  Having little data on the IR properties
of these PNe, it is not possible to estimate on a per-object basis if a
particular PN should be detectable at the sensitivity limits of the survey.
An aperture correction was applied to determine the point source magnitude
in the standard IRAC calibration which used a 12\farcs2 aperture
\citep{reach05}.  Using a smaller aperture can lead to an underestimate
of the nebular flux if there is significant emission outside of the
aperture.  However, an examination of the mosaics shows that almost
all of the IRAC sources are compact and indistinguishable from
other point sources in the field, so it is likely that the photometry
presented here accurately represents the total emission from these
objects.  We have also compared our photometry for a sample of objects
with photometry from the SAGE point source Catalog, which used a
PSF-fitting photometry technique, and find good accord between the two
different extractions, within 0.1-0.2 mag.  Possible reasons for
differences are that the SAGE catalog available at the time was made from
the epoch 1 data only, and that if a PN is slightly extended at the IRAC
resolution, it could result in fitting errors due to a poor match to the
PSF.

\begin{deluxetable}{ccccc}


\tablecaption{Detection Statistics in Each Band\label{tbl-3}}
\tablehead{\colhead{Band} & \colhead{PNe Detected} & \colhead{Median Mag} & \colhead{Std err of median} & \colhead{5 $\sigma$ sens.\tablenotemark{a}}} 
\startdata
J      &      39    &  15.56    &    0.37 & 17.2\\
H      &      36    &  15.30    &    0.32 & 16.2 \\
K      &      42    &  14.34    &    0.27 & 15.6 \\
3.6     &    188    &  15.59    &    0.18 & 19.3 \\
4.5      &   197    &  15.12    &    0.17 & 18.5 \\
5.8     &    126    &  13.33    &    0.22 & 16.1 \\
8.0    &     157   &   12.27    &    0.20 & 15.4 \\
24     &     161   &   7.54     &    0.20 & 10.4  \\
70     &     18     &  2.68     &    0.67 & 3.5 \\
160    &      0     &  \nodata & \nodata &  -0.6 \\
\enddata

\tablenotetext{a}{The 5-$\sigma$ point source sensitivity limits for the IRAC
and MIPS data are from \citet{meixner06}, the 2MASS limits are from the 2MASS
Explanatory Supplement\citep{cutri03}.}



\end{deluxetable}
\subsection{MIPS and 2MASS data}

The \citet{leisy97} positions were used to find MIPS sources
from the SAGE point source catalog \citep{meixner06}.
We found 161, 18, and 0
matching sources in the 24, 70, and 160 $\mu$m bands of the catalog,
respectively.  There were 109 objects detected in all four IRAC bands 
plus MIPS 25 $\mu$m. The large number of 24 $\mu$m detections must represent a
combination of the intrinsically high sensitivity of this MIPS array, and
the peak of the SED for many PNe corresponding to the characteristic
temperature of $\sim$100\,K for dust inside the ionized zone due to
resonantly trapped Ly$\alpha$ photons. The 2MASS sources were also merged
into the SAGE catalog, and those data were extracted along with the MIPS
photometry from the SAGE catalog. There were 39 sources in the SAGE
catalog that had corresponding 2MASS K magnitudes, and slightly fewer
with J and H magnitudes.

\begin{figure*} 
\begin{center}
\includegraphics[angle=0, width=6.0in]{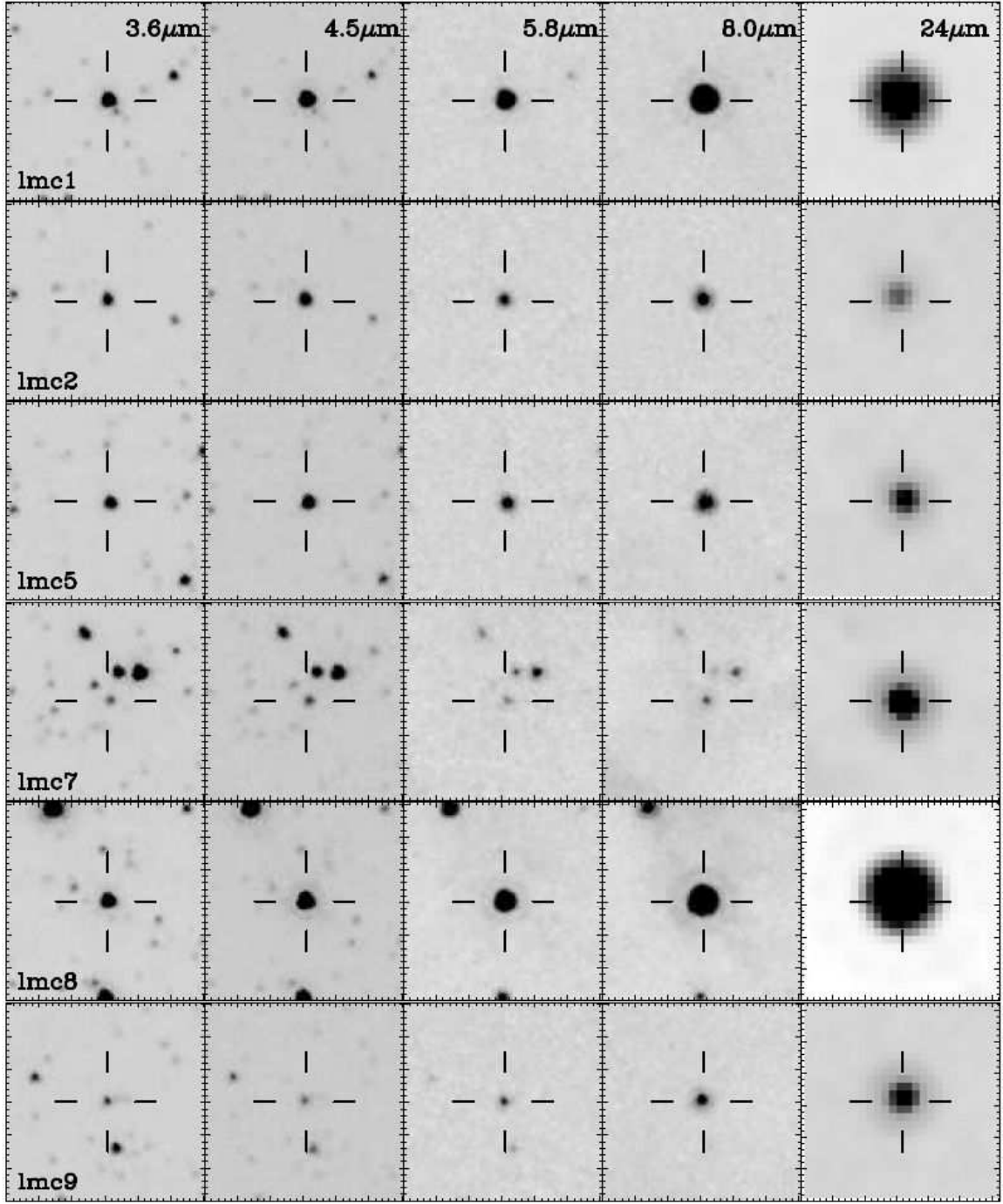}
\end{center}
\caption{IRAC 3.6, 4.5, 5.8, and 8.0 $\mu$m and MIPS 24 $\mu$m images of
nebulae imaged in the SAGE survey.  Each row is labeled with the object
name, and each column shows the image of that object in the different
bands, labeled at the top of the figure.  The images are shown with an
inverse linear grayscale (brighter is darker) which is offset and scaled
in each image to enhance low values near the noise level to better show
faint extended emission.  The images are one arcmin on a side. The crosshair
lines mark the center of the field; the lines 
have a length of 6\arcsec, and are separated by 18\arcsec. This figure shows the first six objects in the
survey list; figures showing all of the objects plus the FITS images of
these mosaics are available at a link listed with this paper at
http://www.cfa.harvard.edu/irac/publications. \label{figmosaics1}}
\end{figure*}

\section{Results}

\begin{figure*} 
\begin{center}
\includegraphics[angle=0, scale=0.95]{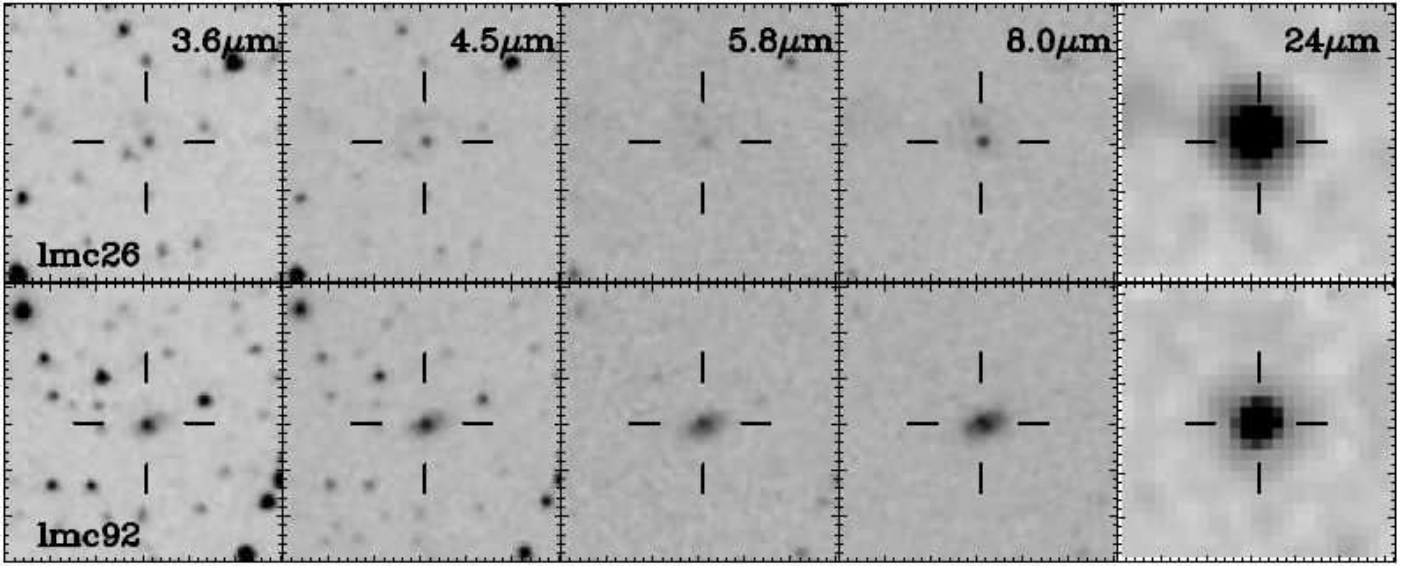}
\end{center}
\caption{Same as Figure \ref{figmosaics1}, except two PNe are plotted that
show signs of extended emission.\label{figmosaics2}} 
\end{figure*}

The IRAC and MIPS 24 $\mu$m images for representative sources are
presented in Figures \ref{figmosaics1} -- \ref{figmosaics2}. A summary of
the number of detections in each of the survey bands and statistics on
their distribution (plus 2MASS) is given in Table \ref{tbl-3}.  For the
2MASS and MIPS 70 and 160 $\mu$m bands, the expected fluxes of most of the
catalog PNe are below the flux limit of the survey and are undetected.  
In other cases, the sources might have been detected except for confusion
with other sources in the field or higher background due to extended in
star forming regions, for example, which will affect the statistics in
those bands. The photometry results are given in Table \ref{tbl-2}.
Objects are labeled in this paper as LMC n, where n is the position in the
\citet{leisy97} catalog, from 1 to 277, and listed in column 1 of Table
\ref{tbl-2}.  The second column lists the distance in arcsec 
between the Leisy et al. 
catalog position and the IRAC source position, which was determined from the 
shortest wavelength band in which the object was detected.  The median distance
is 0.71 arcsec.  Columns 3 and 4
give the position determined from the IRAC images.  Columns 5 -- 10 list the
fluxes determined at each of the Spitzer wavelengths where the object was 
detected.  Column 11 gives the other catalog names of the object listed by
\citet{leisy97}.  Column 12 gives the characteristics of the source 
and field near the nebula in each band, according to the following code:
A=well defined, isolated point source, B=blended with other nearby point
source, C=complex background or distribution of many nearby point sources,
E=extended source, N=no source visible or too faint to determine whether
extended or pointlike. These categories are subjective and were assigned by
two of the authors independently. Their ratings were merged by evaluating 
in more detail any differences. They are meant as a rough guide to the expected
quality of the photometry.

Images of some representative objects\footnote{The full set of images
including FITS files of the IRAC and MIPS mosaics are available at a link
listed with this paper at http://www.cfa.harvard.edu/irac/publications} in
the survey are shown in Figure \ref{figmosaics1}.  The images are
presented as 1 arcmin square inverse grayscale (brighter is darker) images
on a linear scale, and are centered on the \citet{leisy97} catalog
position, indicated by the black hash marks on the images. A blank image is
shown for a particular object only if it
was not in the survey image at a particular band (most commonly due to 
its proximity to the edge of the survey field). Most
of the PNe are unresolved in the IRAC and MIPS images, as the images in
Figure \ref{figmosaics1} indicate.  
Figure \ref{figmosaics2} shows two objects
with extended structure --  LMC 26
and LMC 92. Due to the sometimes crowded
fields and faintness of the extended emission, some of what appears to be
emission from the nebulae could be from superposition of foreground or
background sources unrelated to the PN.  However, in the cases of
LMC 26 and LMC 92, the extended emission is consistent with that
observed with higher resolution optical imaging, so for those sources we
have additional confidence that the IR emission is also associated with
the nebula. Given the resolution of IRAC, any visible extension suggests
an extent of at least 1\farcs2 or 0.3 pc.

\begin{figure*}
\begin{center}
\includegraphics[angle=0,scale=0.6]{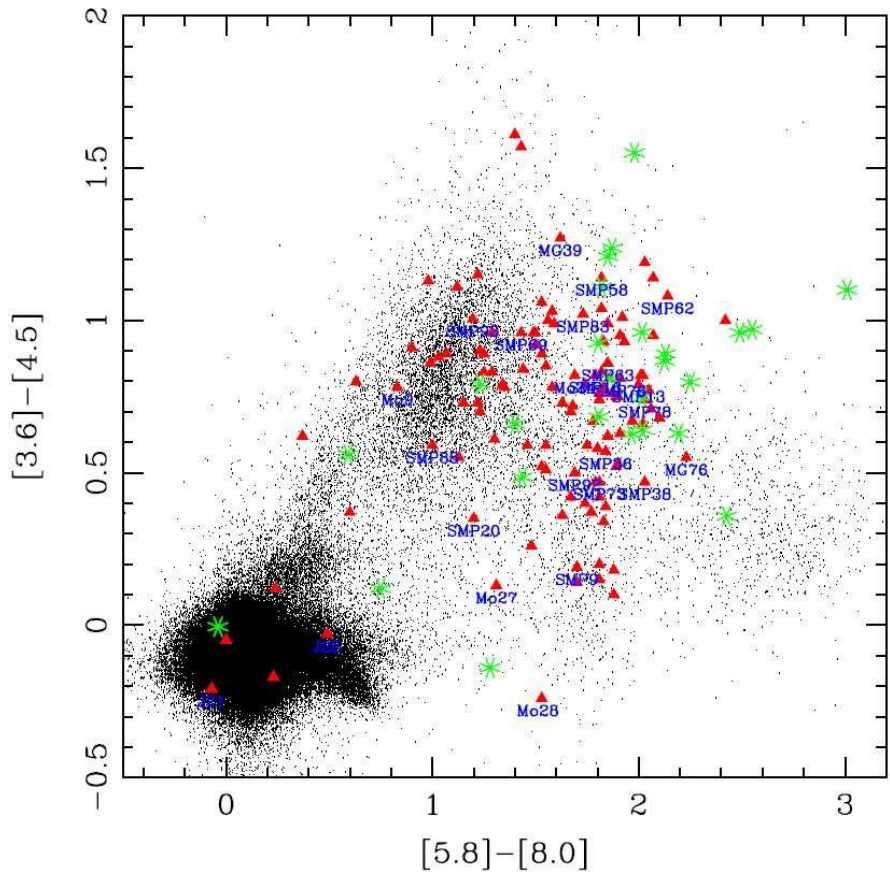}
\end{center}
\caption{ The [3.6]-[4.5] versus [5.8]-[8.0] 
color-color diagram for the LMC PNe and
SAGE catalog sources. The LMC sample from the SAGE data in Table
\ref{tbl-2} is plotted as red triangles, and a subset of these are labeled
directly below the triangles
with blue letters. The subset was chosen to contain examples of different
mid-IR spectral types and optically-determined morphologies. Also plotted
in green are data from several Galactic PNe that have
been observed previously with IRAC \citep[][see the data in Table
\ref{tbl-1}]{hora05,kwok07}. The underlying black
points are a subsample of the SAGE database that have detections at both
3.6 and 8.0 $\mu$m.  All points are plotted in each diagram if the data 
for those bands are available.\label{f12v34}
}
\end{figure*}
\vskip 0.2in
\subsection{Color-Color and Color-Magnitude Diagrams}

\begin{figure*}
\begin{center}
\includegraphics[angle=0,scale=0.6]{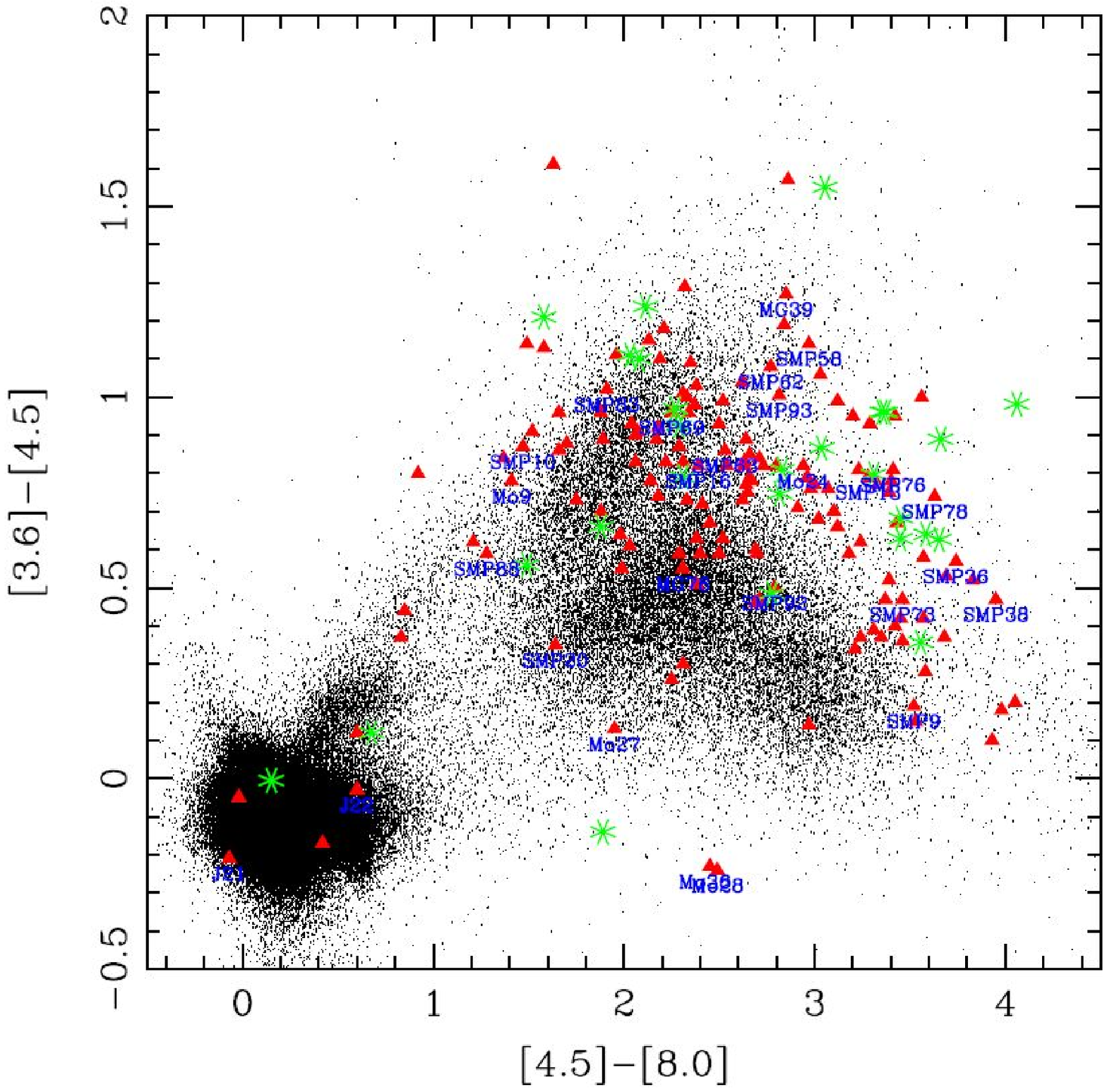}
\end{center}
\caption{ Same as Fig. \ref{f12v34}, except the  [3.6]-[4.5]
versus [4.5]-[8.0]  color-color diagram is plotted.
\label{f1m2v2m4} }
\end{figure*}

Figures \ref{f12v34} -- \ref{fig8vjm8} show various color-color and
color-magnitude diagrams of the datasets.  In all the diagrams, the LMC
sample from the SAGE data in Table \ref{tbl-2} is plotted as red
triangles, a subset of which are labeled with blue letters (not all named
objects are labeled so the figure remains legible).  Also plotted in the
IRAC-only diagrams are data from several Galactic PNe that have been
observed previously with IRAC \citep{hora05,kwok07}.  The photometric data for
these are given in Table \ref{tbl-1}.  The 
underlying black points are a subsample of the SAGE catalog that have
detections at least in both the 3.6 and 8.0 $\mu$m IRAC bands.  Of course,
to be plotted in the various diagrams, the points in addition have to be
detected in all of the bands being plotted in the particular diagram. All
points are plotted in each diagram if the data for those bands are
available.

\subsubsection{[3.6] - [4.5] versus [5.8] - [8.0]}

Figure \ref{f12v34} shows the four-band IRAC color-color diagram.  Most
SAGE sources are stars which are clustered near zero color.  A broad
cluster of points are centered near (1,1), and there is another sparser
group extending from [5.8] - [8.0] colors of 0 to 3, and between [3.6] -
[4.5] colors of 0 to 0.5.  This is similar to the distribution of sources
seen in large shallow surveys such as the IRAC Shallow Survey
\citep{eisenhardt04} where they were shown to be due to galaxies and AGN
\citep{stern05}.  However, in the LMC there are many star-forming regions
and one finds many young stellar objects \citep[YSOs; see ][]{whitney07}
that also populate these areas of the color-color diagram, their position
depending on the age of the YSO and conditions in the circumstellar disks.  
The LMC PNe are scattered in a region centered roughly at (1.5, 0.75), but
extend all the way from objects with zero color to points near the edge of
the region plotted.

The objects near (0,0) (e.g., J21 and J22) are likely dominated at these
wavelengths by light from their central star. One does not expect to see
many PNe near zero colors in this particular plot because of the brighter
detection threshold at 8 $\mu$m compared with the other IRAC bands.  
However, none of the Galactic PNe are near this point, and any PN with
zero colors is likely to be a highly evolved object in which the nebular
gas is recombining so that other processes that are seen in the MIR are
absent. These ``zero color'' objects appear near the main locus of stars
in this and the following diagrams, except for those including wavelengths
of 24 $\mu$m and beyond. Another possibility is that since several of
these are in crowded regions, they could be misidentified and the
photometry is for a star that is near the PN in the image.

The majority of the survey objects appear red in one or both of these
colors which is likely due to either line emission from ionized gas, PAH
band emission, or continuum emission from dust.  In several of the objects
where mid-IR spectra are available \citep[for example, those in the
Spitzer GTO observations of LMC PNe; see][]{salas04, salas05, salas06}, we
see that some PNe have all three of these components contributing to their
emission.

\subsubsection{[3.6] - [4.5] versus [4.5] - [8.0]}

In the four-band IRAC color-color diagram in Figure \ref{f12v34}, most of
the PNe are in the range of [5.8] - [8.0] colors of 1-2.  Since both the
5.8 and 8 $\mu$m bands include emission from PAH features, and the
strengths of the features are well correlated \citep[][their Fig.  
18]{cohen89}, one might expect the PAH features to have little effect when
comparing the strengths of these bands.  Also, if the PN has significant
warm dust continuum, it will be detected in both the 5.8 and 8.0 bands. In
the [3.6] - [4.5] vs. [4.5] - [8.0] diagram shown in Figure
\ref{f1m2v2m4}, the [4.5] - [8.0] color of the PNe have a greater spread,
spanning the range $\sim$ 1 -- 4.  This could be in part due to the lack
of PAH features and the much fainter emission from warm dust in the 4.5
$\mu$m band, enhancing the [4.5] - [8.0] color.  In fact, one sees that
several PNe with strong continuum emission from warm dust such as Hb 5, Hb
12, and SMP 76 appear at the extreme right of the PNe distribution,
whereas objects with emission line spectrum and little or no dust
continuum such as NGC 2440, NGC 246, and SMP83 appear on the right side of
the main PNe group near (2,0.75).  PNe that have strong PAH and forbidden
line emission in their 5 - 15 $\mu$m spectra such as SMP 36 and SMP 38
\citep{salas05} appear on the right side but lower in the [3.6]-[4.5]
color than PNe with continuum-only (in the 3-10 $\mu$m range) such as SMP
62. This bluer [3.6]-[4.5] color might reflect the contributions of the 3.3
$\mu$m PAH band and/or H recombination lines such as Pf$\gamma$ to the 3.6
$\mu$m band.

\begin{figure*}
\plottwo{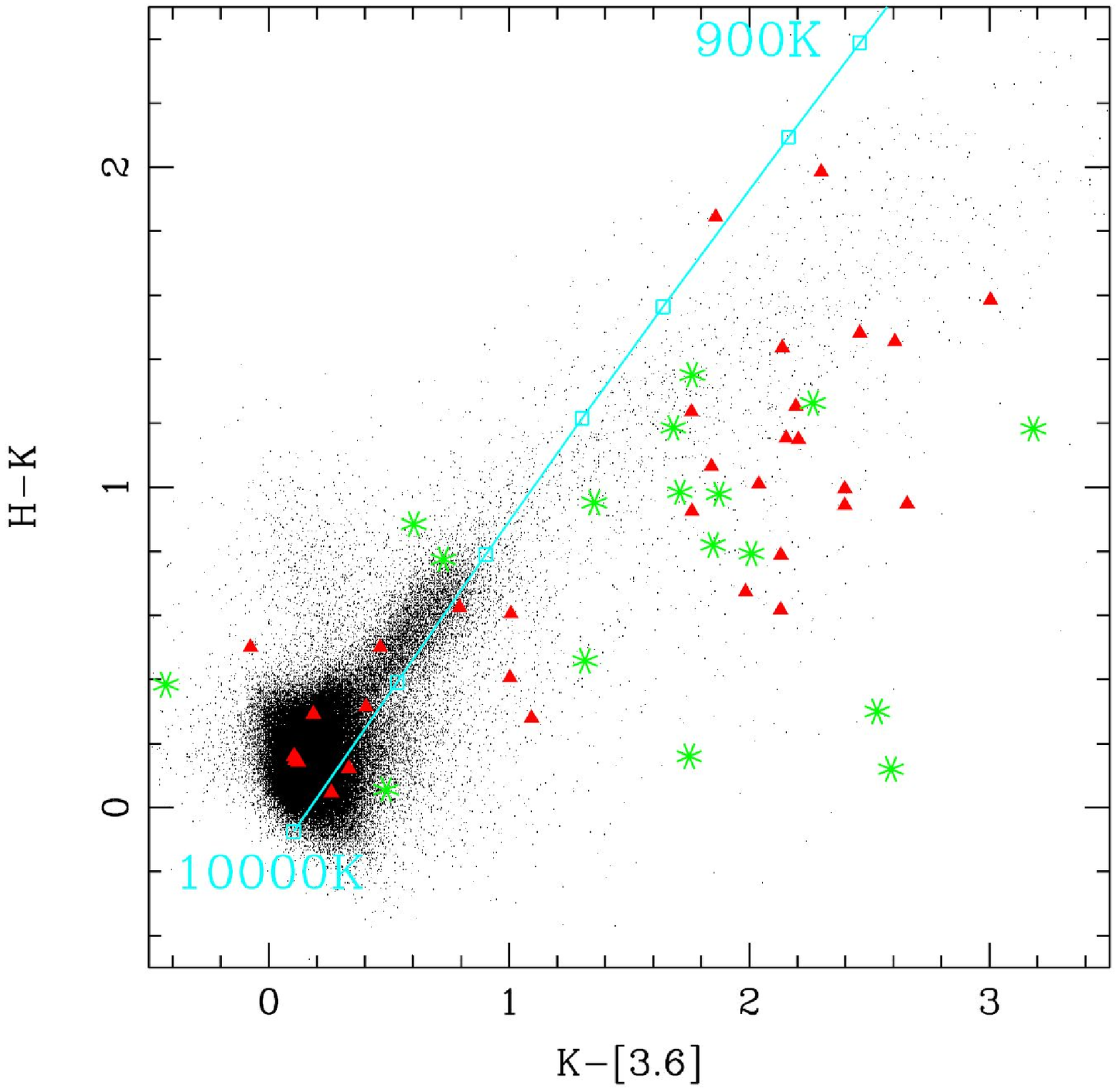}{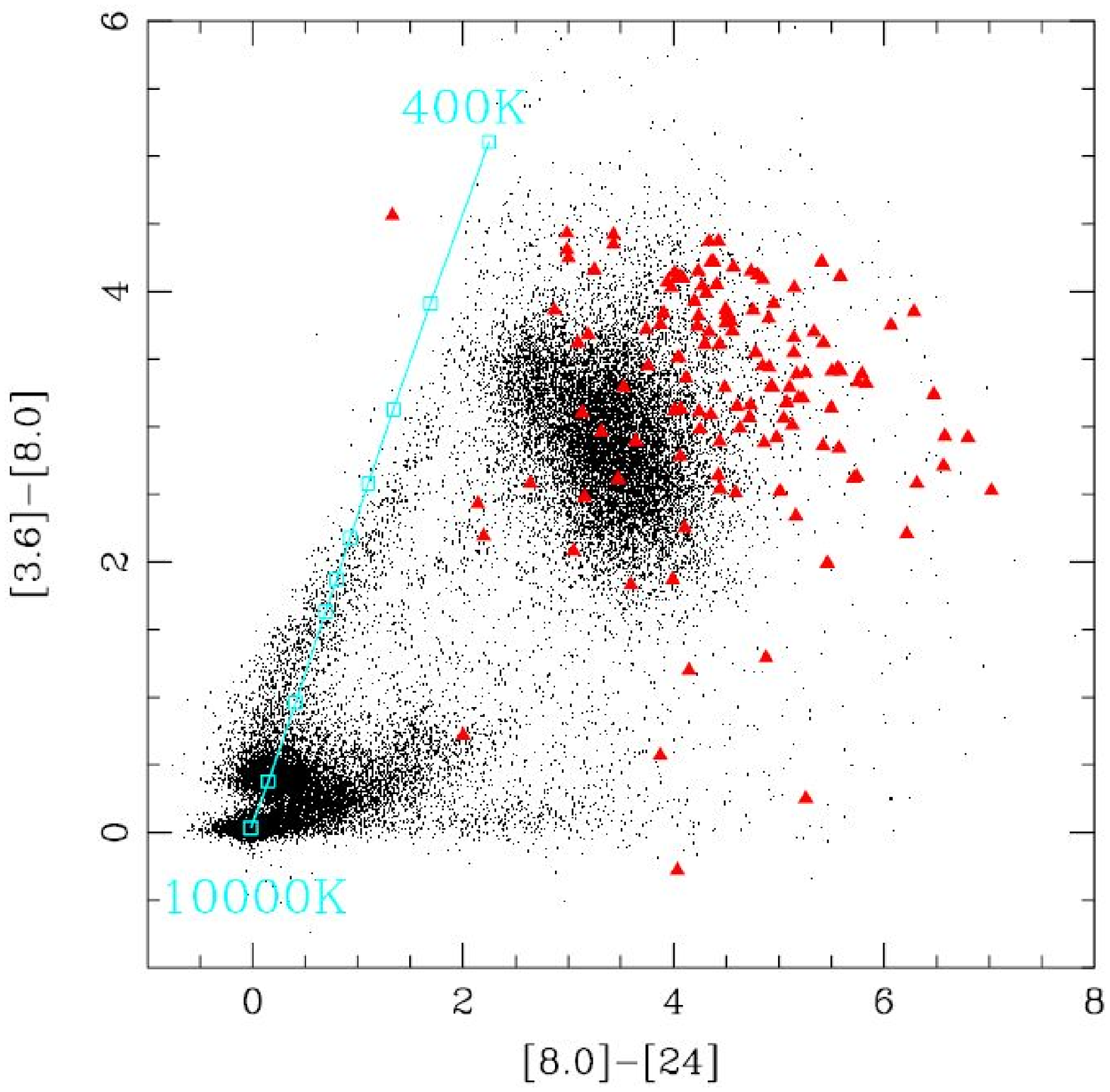}

\caption{Same as Fig. \ref{f12v34}, except the H - K versus K -[3.6] ({\it
left}) and [3.6]-[8.0] versus [8.0]-[24] ({\it right}) color-color
diagrams for the LMC PNe and SAGE sources. \label{fhmkvkm1} The colors of
a blackbody of various temperatures are plotted in each figure as cyan
squares connected by a line.  In the figure on the left, temperatures of
10,000K, 3000K, 2000K, 1500K, 1250K, 1000K, 900K (from lower left to upper
right) are plotted.  In the figure on the right, temperatures of 10,000K,
3000K, 1500K, 1000K, 900K, 800K, 700K, 600K, 500K, 400K (from lower left
to upper right) are plotted. \label{f1m4v4m24} } 
\end{figure*}

\subsubsection{Near-IR to Mid-IR and Mid-IR to Far-IR color-color diagrams}

In the H - K versus K - [3.6] diagram in Figure \ref{fhmkvkm1}, the PNe
separate into two main groups, one near the locus of most of the stars
near (0.2, 0.2), and the other clustered around the area near (2.2, 1.0),
with a large gap in the K - [3.6] color between 1 and 2. In the first
group, there is perhaps evidence for a subgroup that is slightly redder
than the PNe clustered around the stellar locus at around (1.0, 0.5) in
the diagram. These include SMP 62 for which the IRS spectrum shows does
not show strong PAH emission, only cool dust continuum and some forbidden
line emission at wavelengths longer than the IRAC bands \citep{salas05}.

The colors of blackbody emission at various temperatures are plotted on
the graph.  Colors as red as (2.2,1.0) are characteristic of thermal
emission from $\sim$800K dust (\citet{allen74}, their Fig. 2; 
\citet{cohen79}, their Fig. 12).  Even colors of (1.0,0.5) can be caused
by this mechanism although these require higher temperature grains,
$\sim$1000K.  An offset between the latter PNe and the group that is
star-dominated can be understood if the materials around the PNe have
condensation temperatures just above 1000K so that no hotter grains exist.  
Both SMP62 and SMP88, which have this putative hot dust, are dense nebulae
($N_e \sim 7000$ cm$^{-3}$: Reid 2007, private communication), consistent
with local condensation of grains in clumps.

\begin{figure*}
\plottwo{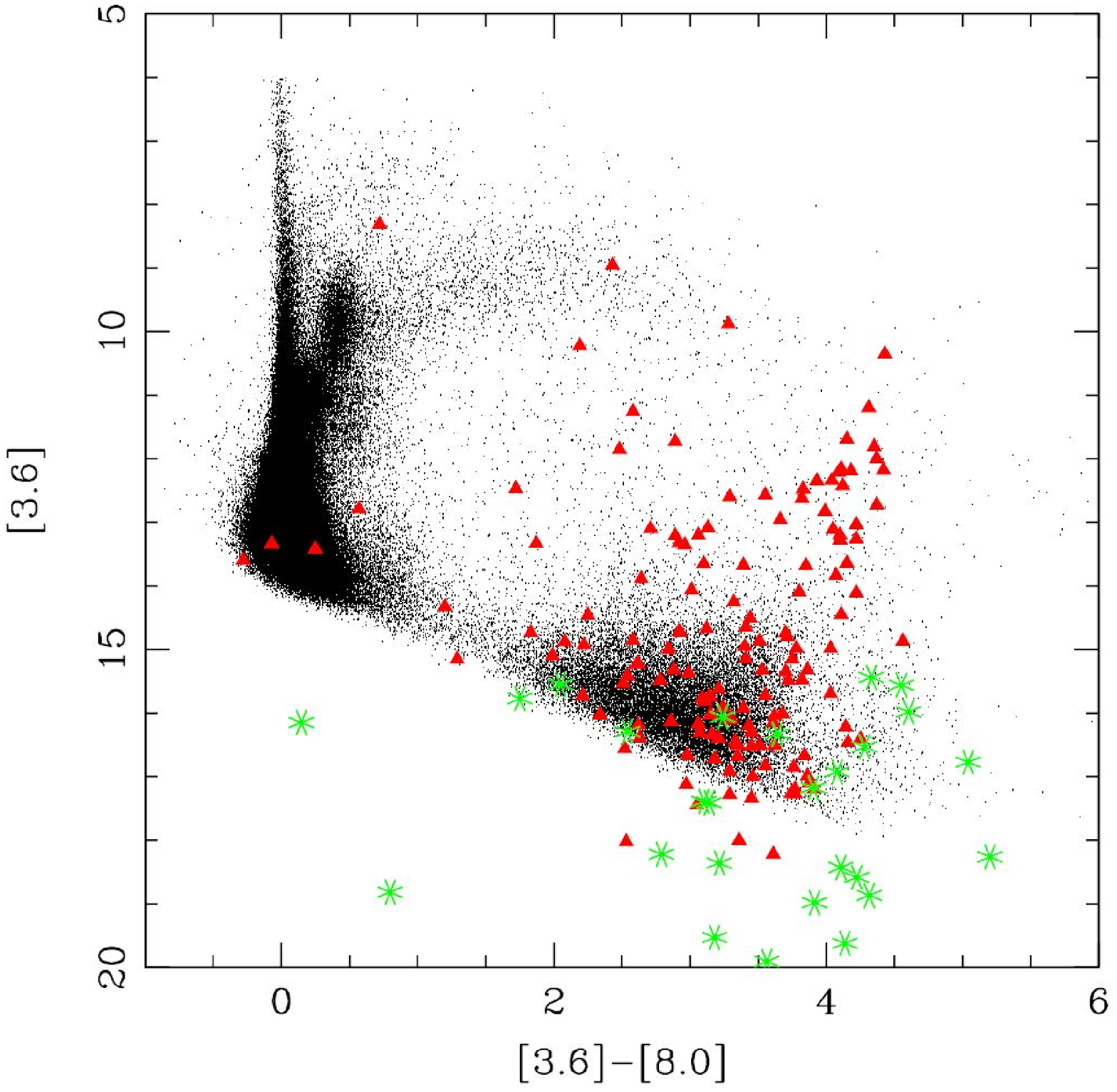}{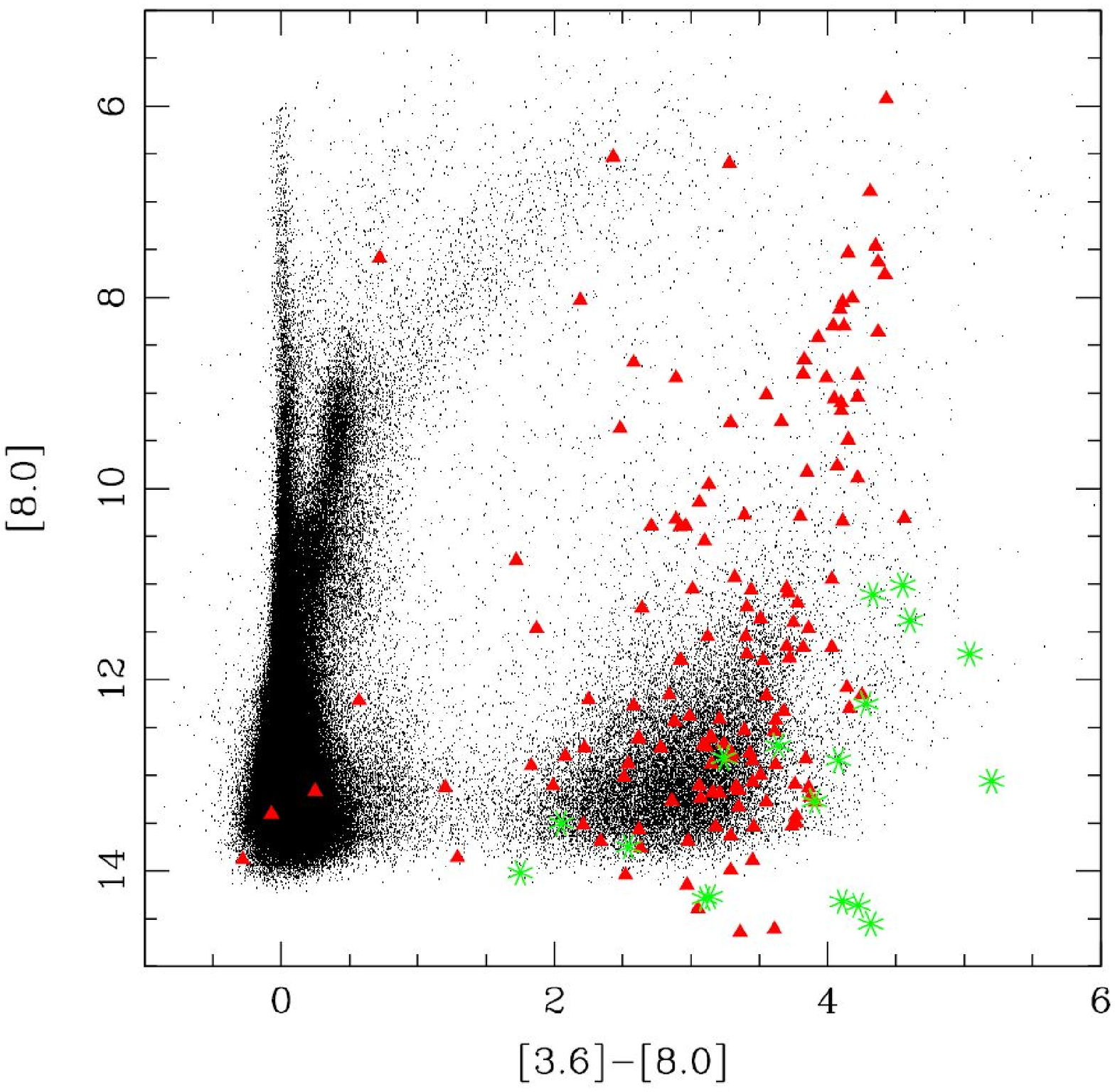}
\caption{Left: Same as Fig. \ref{f12v34}, except the IRAC [3.6] versus
[3.6]-[8.0] ({\it left}) and \label{fig3v3m8} [8.0] versus [3.6]-[8.0]
({\it right}) color-magnitude diagrams are shown. The magnitudes of the
Galactic PNe were shifted to the value they would have at the distance of
the LMC (see text).\label{fig4v1m4}}
\end{figure*}

Figure \ref{f1m4v4m24} also shows the [3.6] - [8.0] vs. [8.0] - [24]
color-color diagram. The colors of a blackbody at temperatures from
10,000K to 400K (lower left to upper right) are plotted as cyan points
with a line through them. The PN population is well-separated from the
distribution of main sequence stars which are clustered near the origin,
and the cluster near 3000K which are likely red giant stars
\citep{meixner06}.  There is a line of cooler objects extending up the
blackbody curve that is more evident here because of the larger sample of
objects included in the plot.  The center of the PN distribution is
located above and to the right of the large number of objects centered
roughly at (3,3), which are possibly extragalactic objects or
YSOs \citep{meixner06}.

\subsubsection{The [3.6] - [8.0] color-magnitude diagrams}

Figure \ref{fig3v3m8} shows the [3.6] vs. [3.6] - [8.0] and the [8.0] versus [3.6] - [8.0] color-magnitude
diagrams.  These and the following color-magnitude diagrams show the wide
range of luminosities of the PNe in the sample. The Galactic sample are 
plotted by adjusting their magnitudes to place them at the distance of
the LMC, assuming a distance modulus of 18.5 for the LMC, and the PNe 
distances as given in the Strasbourg PNe catalog \citep{acker92}. 
In the [3.6] versus [3.6] - [8.0] diagram, most of the LMC PNe are
roughly grouped around the position (3.5, 16). They have similar 
[3.6] - [8.0] colors as the Galactic sample, but the sensitivity cutoff of
the SAGE survey passes roughly through the middle of the Galactic PNe
distribution, indicating that we are likely not detecting a large fraction
of the LMC PNe. 
In both of these diagrams, there are again a few PNe
appearing near the distribution of LMC stars on the left side of the
diagram.  There are also a few bright objects that appear along the
top of the diagram, with colors similar to the objects classified by \citet{blum06} as
being ``Extreme AGB'' stars.  One would expect that PNe could not be confused 
with extreme dust-shrouded AGB stars because there would be no optical spectroscopy 
to confirm their status as PNe.  However, SMP11 is clearly an optical PN yet its
MIR spectrum suggests that it is a post-AGB object \citep{salas06}.
Other MIR-bright PNe could likewise be in transition between post-AGB and proto-PN
stages.  \S4 contains further discussion of these bright PNe.
Most of the LMC PNe are on the right side of the diagram, 
near the clump of what are
probably background galaxies in the SAGE catalog \citep[see][Figure
5]{blum06}.  Also, the distribution of PNe are similar to the colors
expected for YSOs \citep{whitney07}.

\subsubsection{Other color-color and color-magnitude diagrams}

\begin{figure*}
\plottwo{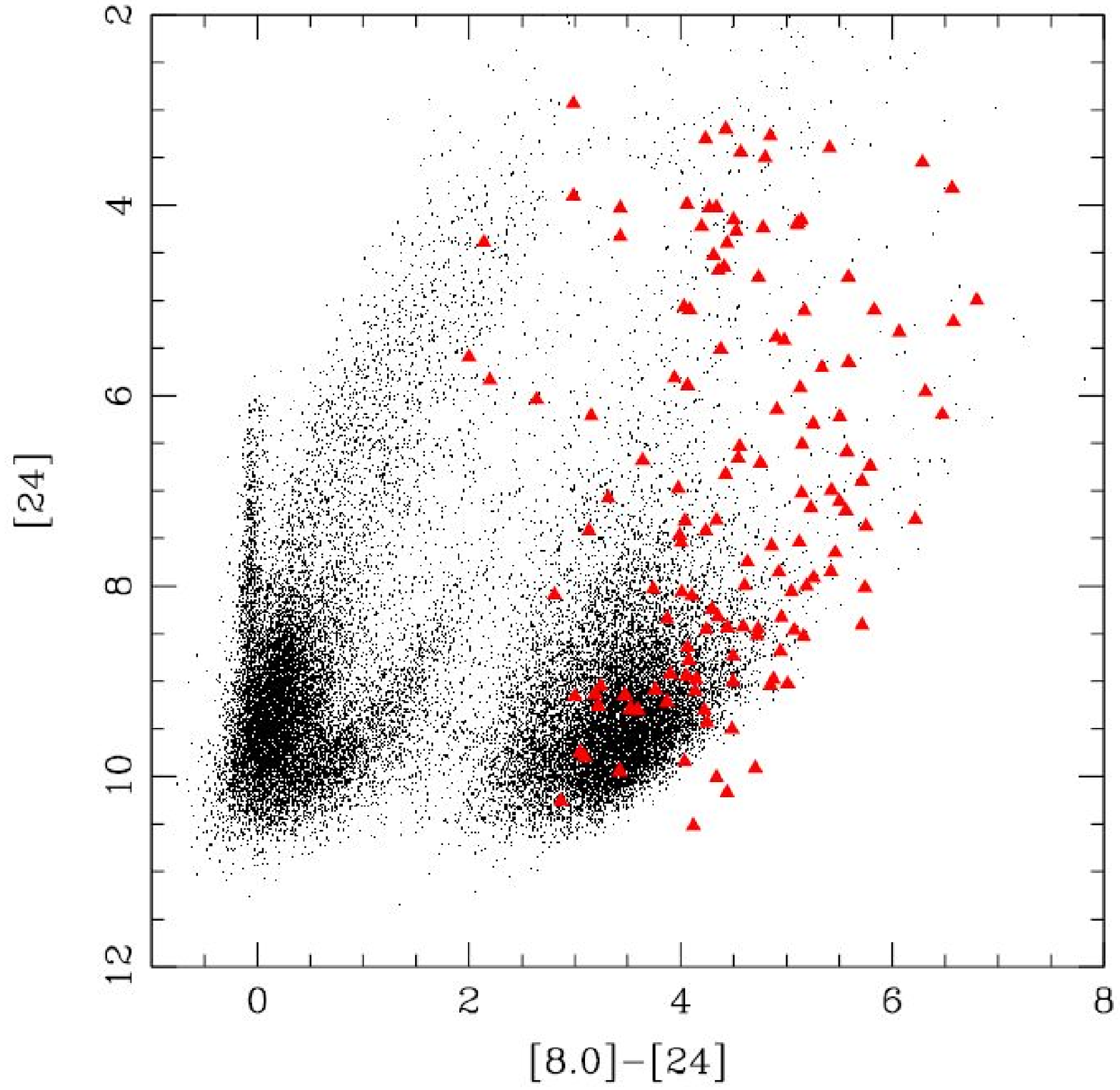}{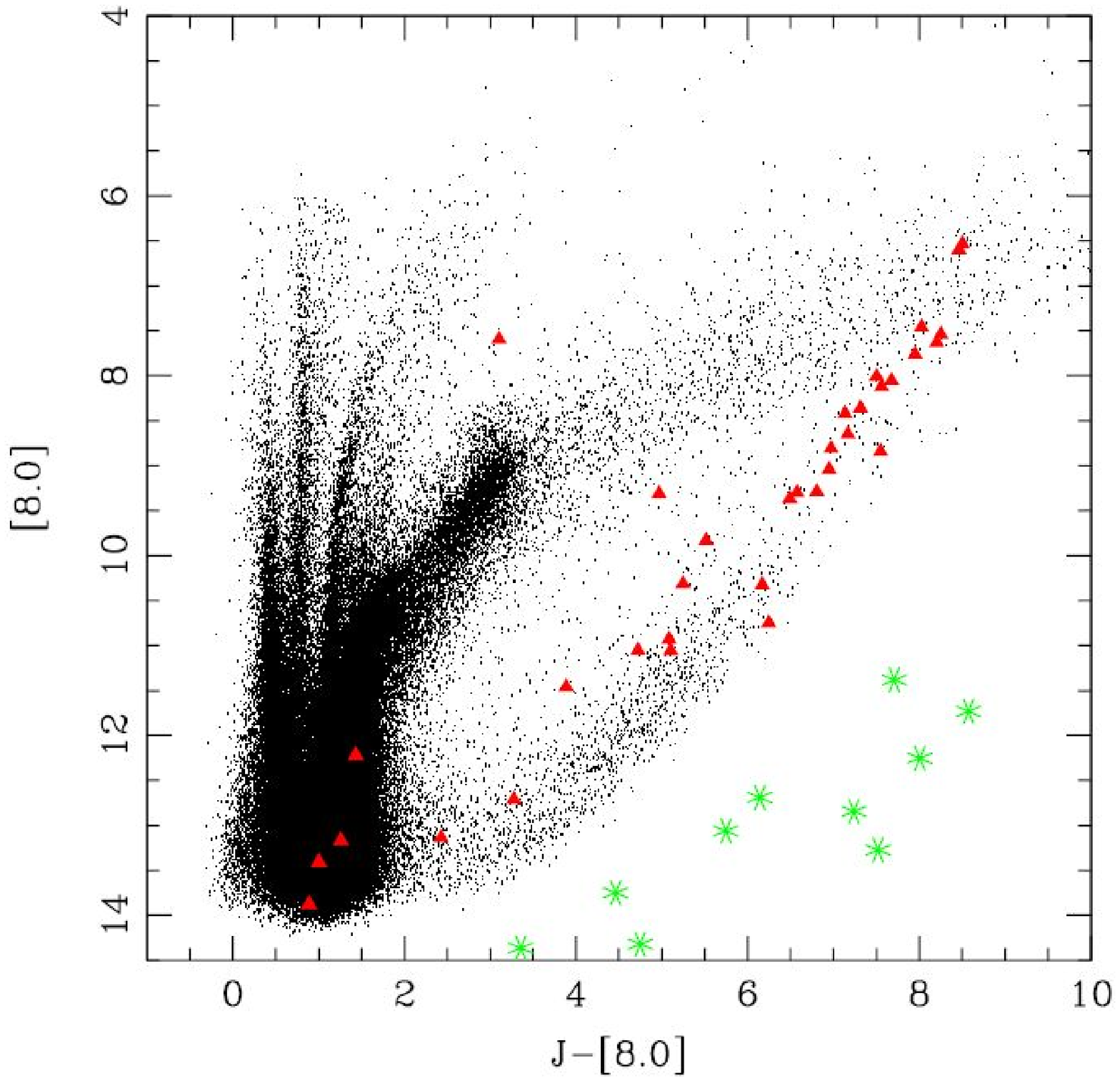}

\caption{Same as Fig. \ref{fig4v1m4}, except the [24] versus [8.0]-[24]
({\it left}) and \label{fig8vjm8} [8.0] versus J - [8.0] ({\it right})
color-magnitude diagrams are shown. \label{fig24v8m24}}

\end{figure*}

Figure \ref{fig4v1m4} shows the [8.0] versus [3.6] - [8.0] diagram, 
and Figure \ref{fig24v8m24} shows the [24] versus [8.0] - [24] diagram.  
These figures can be compared to Figures 5 and 6 of \citet{blum06}, which
labeled a few bright PNe which appeared at [8.0] - [24] colors of 3 - 5,
and [24] of 1 - 2.  From our Figure \ref{fig24v8m24}, it can be seen that
the majority of PNe are in this color range, but are detected here at
fainter magnitudes. There is significant overlap between the PNe and the
large group of objects labeled ``galaxy candidate'' and ``no 2MASS J'' in
the \citet{blum06} figure.  However, the distribution of PNe is centered
at a position brighter and slightly redder than the galaxy candidate
distribution.

Figure \ref{fig8vjm8} shows the [8.0] versus J - [8.0] diagram, which can
be compared to Figure 4 of \citet{blum06}. There is a set of PNe that fall
slightly above the group of objects identified as galaxy candidates in
Blum et al., but a larger group of objects appear at the red tip of that
distribution, slightly fainter than the extreme AGB star distribution.  
The objects that have strong PAH and continuum dust emission, such as SMP
38 and SMP 76, appear on the red end of the distribution, whereas objects
with strong ionized gas lines such as SMP 62 appear less red. This
distribution is possibly due to the 6$-$9 $\mu$m plateau emission and the
several PAH bands in the 8.0 $\mu$m IRAC band making objects like SMP 38
appear more red \citep{salas05}. These strong PAH features are typical of
young PNe, for example NGC 7027 \citep{salas01}.  In more evolved PNe, the H recombination 
lines dominate the near-IR emission \citep{hora99}. For example, the 
brightest near-IR line in the J band is 
Pa$\beta$ line at 1.28 $\mu$m. These lines could make the objects appear
bluer than they would from the stellar continuum alone.
The Galactic sources are at fainter [8.0] and redder J - [8.0] colors than the 
LMC population, falling below the sensitivity limits of the 2MASS and/or the
IRAC LMC data. The difference in the distributions
could be due to the fact that the Galactic sample 
was chosen with the criteria that they are relatively nearby objects that happen to
be in the GLIMPSE and 2MASS surveys, whereas in the LMC our sensitivity limits
have selected the brightest objects in that galaxy.  We might find that many of the 
sources that are detected by IRAC but missing from the 2MASS database lie in 
the region indicated by the Galactic sources.

\begin{figure}
\begin{center}
\includegraphics[angle=0,scale=0.4]{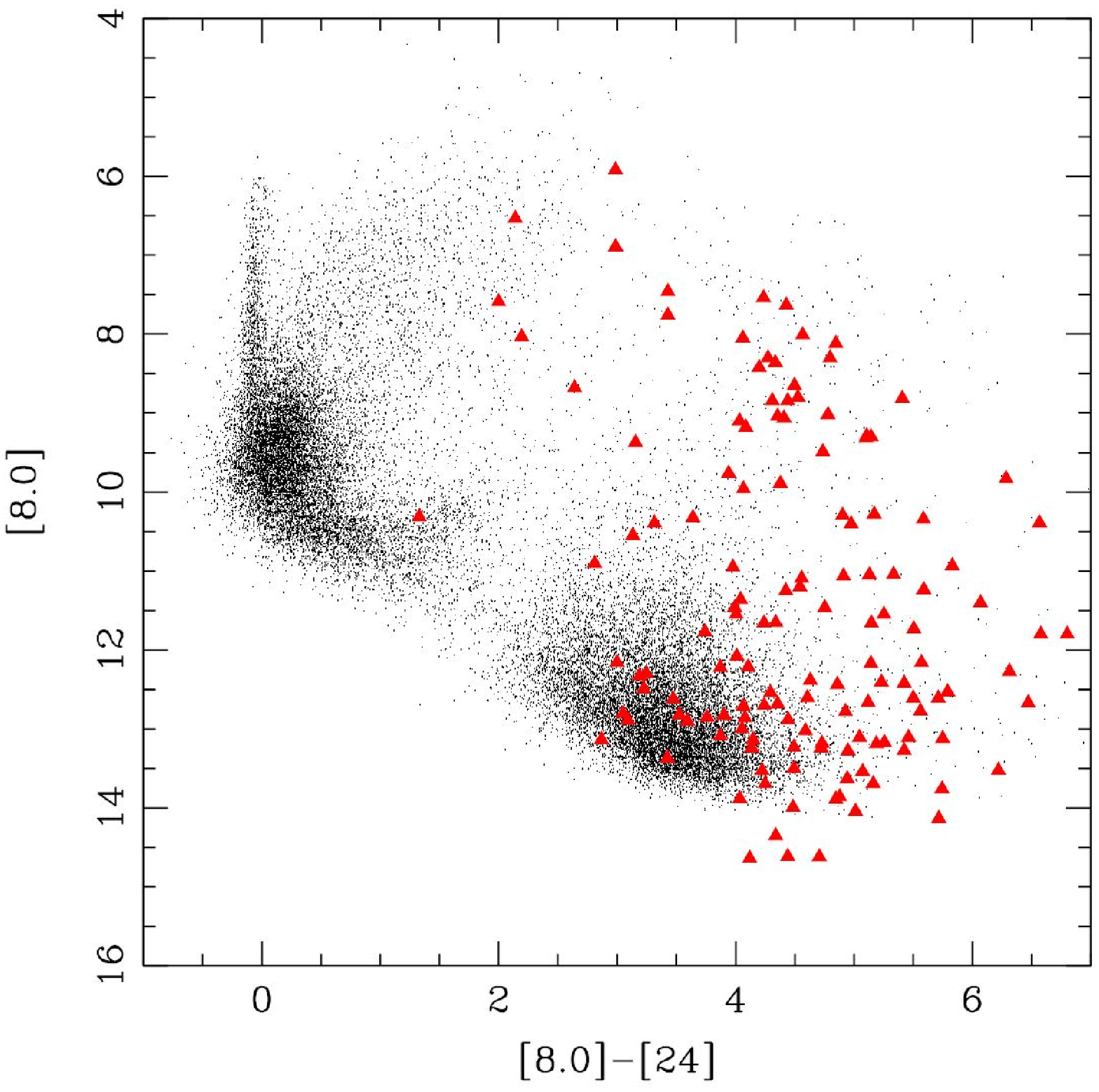}
\end{center}
\caption{Same as Fig. \ref{f12v34}, except the [8] versus [8.0]-[24]
color-magnitude diagram is shown. \label{fig8v8m24}} 
\end{figure}

Figure \ref{fig8v8m24} shows the [8.0] versus [8.0] - [24] color-magnitude
diagram.  The lower left corner of the diagram is blank due to the
sensitivity limit of the survey in the 8.0 and 24 $\mu$m bands.  The
objects near zero color are stars, and a significant population of objects
centered near (3.5, 13) are likely YSOs or extragalactic background
objects \citep{meixner06,whitney07}.  The distribution of PNe overlap with
this region, but should be much less common, so it is likely that most of
the other SAGE catalog objects in this range are YSOs or background
objects.

\begin{deluxetable*}{lccccccc}
\tabletypesize{\scriptsize}
\tablecaption{Galactic Planetary Nebula Magnitudes\label{tbl-1}}
\tablewidth{0pt}
\tablehead{
\colhead{Name} & \colhead{J} & \colhead{H} & \colhead{K} & \colhead{[3.6]} & \colhead{[4.5]} & \colhead{[5.8]} & \colhead{[8.0]}}
\startdata
Hubble 12 & \nodata & \nodata & \nodata & 6.94 & 5.98 & 5.1 & 2.61 \\
NGC 2440 & 10.33\tablenotemark{a} &  10.50 &  9.68  & 7.83 & 7.02 & 6.05 & 4.19 \\
NGC 246 & \nodata & \nodata & \nodata & 8.91 & 7.8 & 7.59 & 5.77 \\
NGC 650 & \nodata & \nodata & \nodata & 9.72 & 8.51 & 8.78 & 6.93 \\
NGC 3132 & 9.71\tablenotemark{a} & 9.70 & 9.54 & 7.79 & 7.13 & 6.65 & 5.25 \\
NGC 6543 & \nodata & \nodata & \nodata & 7.56 & 6.59 & 6.87 & 4.32 \\
Hubble 5 & \nodata & \nodata & \nodata & 7.06 & 6.17 & 4.64 & 2.51 \\
IC 4406 & \nodata & \nodata & 10.1\tablenotemark{b} & 8.89 & 8.1 & 7.02 & 5.79 \\
PNG 002.7-52.4 & \nodata & \nodata & \nodata & 11.03 & 9.93 & 10.86 & 7.85 \\
Mz 1 & \nodata & \nodata & \nodata & 7.27 & 7.41 & 6.8 & 5.52 \\
NGC 2346 & \nodata & \nodata & \nodata & 7.05 & 6.49 & 5.59 & 5 \\
G010.1045+00.7414\tablenotemark{c} &  10.589 &10.181 & 9.196& 7.48 &   5.93&    4.86 &   2.88\\
G011.7469-00.6475 &  11.759 &11.561 &10.298& 8.03 &  7.40 &   5.72  &  3.75\\
G035.5650-00.4910 &  13.019 &11.360 &10.585& 9.86 &   8.93&    8.45 &   6.64\\
G055.5077-00.5558 &  11.582 &11.276 &10.297& 8.42 &  7.79 &   6.54  &  4.34\\
G056.4016-00.9032 &  14.205 &13.712 &12.760& 11.41& 10.66 &   9.86  &  7.84\\
G062.4909-00.2698 &  15.325 &14.584 &14.464& 11.87& 10.64 &   10.40 &  8.52 \\
G300.2818+00.6619 &  10.569 &10.473 &10.417& 9.93 & 9.13  &  8.07  & 5.82 \\
G300.4289-00.9815 &  15.282 &13.648 &13.190& 11.87& 11.39 &  10.05 &  8.61 \\
G302.3730-00.5390 &  14.287 &13.954 &13.654& 11.12& 10.44 & 8.79   & 6.99 \\
G306.4740+00.2590 &  10.500 &10.271 & 9.887& 10.32& 10.20 & 10.27  & 9.52 \\
G315.0266-00.3744 &  15.221 &14.735 &13.552& 10.37&  9.41 &  8.07  & 6.05 \\
G318.9300+00.6930 &  14.988 &13.285 &12.492& 10.48& 10.13 &  9.00  & 6.57 \\
G321.0230-00.6990 &   9.216 &8.722  & 8.375& 10.08& 9.44  & 7.88 &  5.86 \\
G322.4700-00.1800 &  11.805 &11.142 & 9.956& 8.27 & 7.29  & 6.46 &  3.23 \\
G333.9279+00.6858 &  10.306 &9.235  & 8.475& 9.76 & 7.00  &  7.09 &   4.56 \\
G342.7456+00.7521 &  11.121 &9.137  & 8.252& 7.65 & 7.65  &  7.46 &   7.50 \\
G343.9923+00.8355 &  12.292 &11.791 &10.440& 8.68 & 7.81  &  6.90 &  4.77 \\
\enddata
\tablenotetext{a}{JHK data from \citet{whitelock85}, using a 24'' aperture.}
\tablenotetext{b}{K data from \citet{allen74}}
\tablenotetext{c}{Data for all objects starting with ``G'' are from \citet{kwok07}.  Note that the JHK magnitudes
are from the 2MASS point source catalog, so they might be underestimates of the
total flux from the central star plus nebula.}
\end{deluxetable*}

\subsection{Planetary Nebulae Spectral Energy Distributions}

\begin{figure}
\includegraphics[angle=0, scale=0.56]{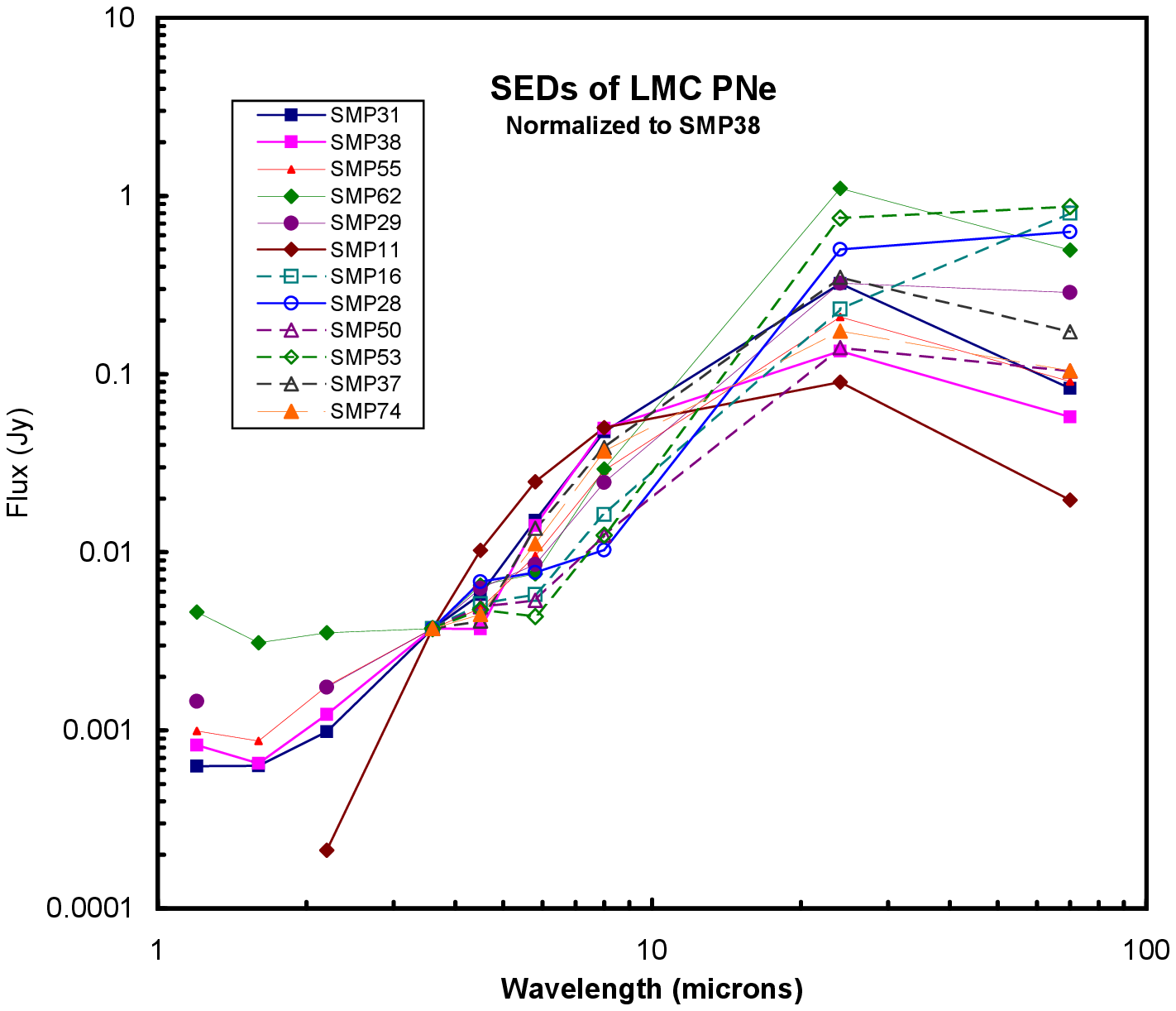}
\includegraphics[angle=0, scale=0.5]{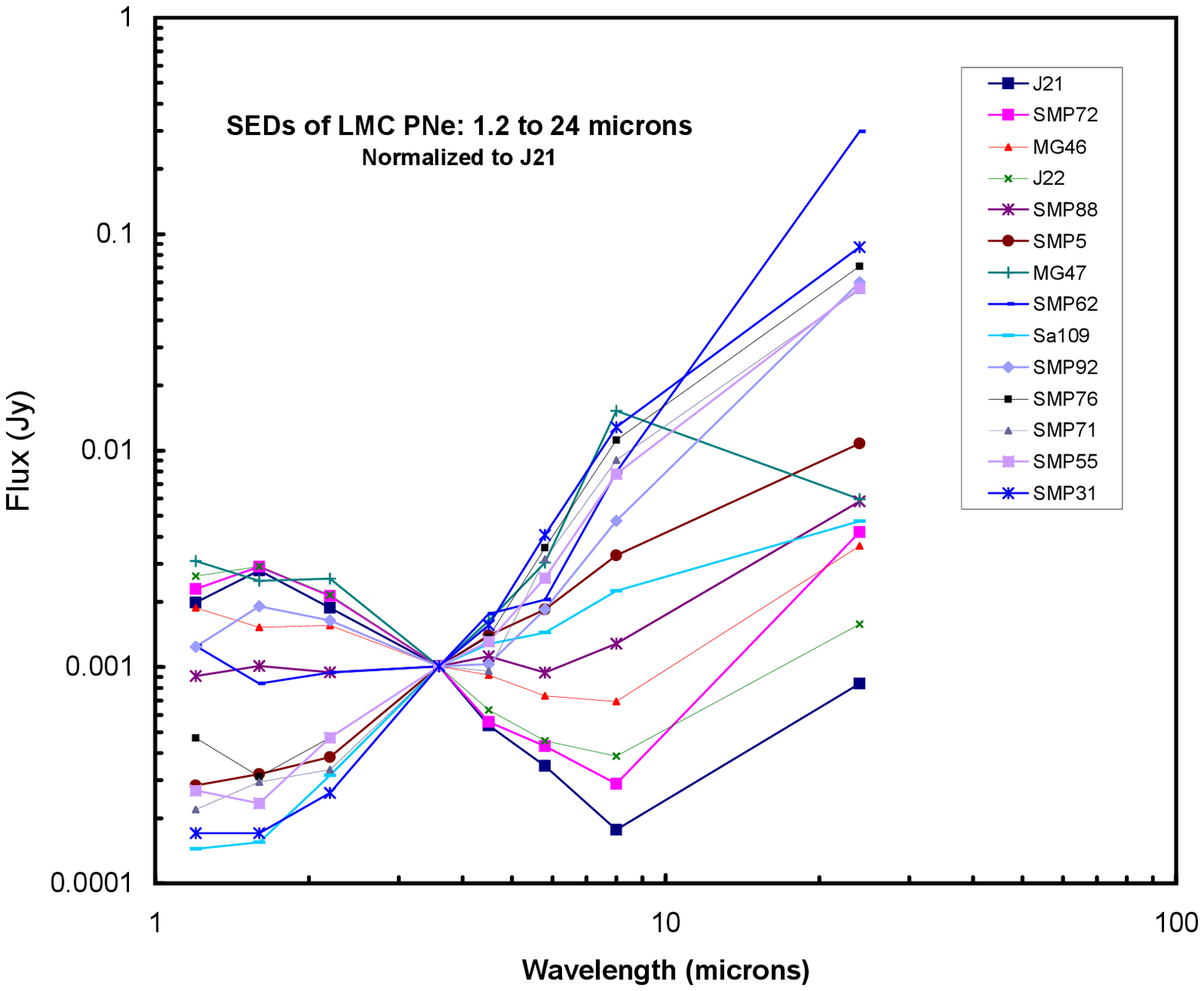}
\caption{Planetary nebulae SEDs. (Top) SEDs from 1.2 - 70 $\mu$m. The SEDs
have been normalized to the 3.6 $\mu$m flux of SMP 38. \label{sed70}
(Bottom) SEDs from 1.2 - 24 $\mu$m.  The SEDs have been normalized to the
3.6 $\mu$m flux of J21. \label{sed24}} 
\end{figure}

Two examples of Spectral Energy Distributions (SEDs) are shown in Figure
\ref{sed70}. The SEDs have been normalized in each plot to the 3.6 $\mu$m
flux of the objects specified in the figure, in order to easily compare
the SED shapes.  In the top plot of Figure \ref{sed70} the SEDs of all of
the PNe that were detected by 2MASS and in the 70 $\mu$m MIPS channel are
plotted, in order to show the widest wavelength range possible in the
dataset.  The sensitivity limit of the 70 $\mu$m data is near the 20 mJy
level of the lowest plotted point. The requirement of a 70 $\mu$m
detection has selected objects with similar SEDs.

In the bottom plot of Figure \ref{sed24}, the 70 $\mu$m requirement was
dropped and a selection of sources were plotted that have 2MASS, IRAC, and
24 $\mu$m detections (there were 29 such sources in total).  
Here the greater diversity of SEDs is seen that is
also reflected in the color-color and color-magnitude diagrams shown
above. With only one exception in this subsample (MG 47), all of the PNe show a
rise between 8-24 $\mu$m indicating a cool dust component.  There were a 
total of 26 sources in this group of objects with 24 $\mu$m flux 
larger than the 8 $\mu$m flux, or roughly 90\%.  The objects with 8 $\mu$m 
larger than the 24 $\mu$m flux may belong to a different class of PNe, 
perhaps having a warmer dust component that dominates the IR luminosity.
The 24 $\mu$m data are not available for the Galactic sources in our sample
to compare to the LMC sample.

The minimum in
the SED varies from 8 $\mu$m down to the J band, and is likely influenced
by the relative strengths of the stellar continuum and scattered light in
the near-IR, H$_2$ emission and warm dust emission in the IRAC bands, PAH
emission that can enhance the 5.8 and 8.0 $\mu$m emission, cooler
dust in the 24 $\mu$m band, and forbidden
line emission from the ionized gas which could affect multiple bands.

\subsection{Individual Planetary Nebulae} 

Figure \ref{figmosaics2} shows two PNe in the catalog that have 
clearly resolved 
extended emission in the IRAC images.  These are discussed in the sections
below. 

\subsubsection{LMC 26 (SMP 27)}

A very large faint halo is visible at a distance of $\sim$ 7\arcsec\ from
the core in the IRAC images. \citet{shaw01} also detected an arc visible
at 6\farcs25 to the northwest in HST images, and hypothesized that the
largest outer arc consists of remnants of the giant branch star that
preceded the nebula. Shaw et al. also detect a compact quadrupolar nebula
around the central star that is too small to be resolved in the IRAC
images.  However the arcs can be seen in the IRAC images to extend much
further around the star than in the HST images. The arcs are traceable at
4.5 $\mu$m as an almost complete ring, particularly on the eastern side of
the central nebula, and more diffuse emission is visible completely
surrounding the object in the 8.0$\mu$m image.

\subsubsection{LMC 92 (SMP 93)}

The PN appears bipolar in the IRAC images, with a bright core and two
distinct lobes extending to the northwest and southeast. The MIPS
resolution is not sufficient to resolve the nebula, but an elongation is
apparent in the 24 $\mu$m image at the same position angle.  The infrared
structure is consistent with that seen by \citet{shaw01} in their HST
image, which they determined to have a maximum extent of 6\farcs4 or 1.57
pc. The lobes are brighter relative to the core at the longer IRAC
wavelengths.

\section{LMC PN Luminosity Function}

PNe in the Magellanic Clouds and in more distant galaxies follow a
universal form of [\ion{O}{3}] luminosity function (LF) which is thought
to arise because [\ion{O}{3}] is the dominant optical coolant
\citep{jacoby89,ciard89}.  The absolute [\ion{O}{3}] magnitude
corresponding to the sharp PN luminosity function (PNLF) cut-off is now
regarded as among the best standard candles for cosmology \citep{jacoby90,
ciard05, jacoby06}.  It would be of interest to know whether LMC PNe
reveal any IR analogs to the optical character of the LFs because a
cosmological distance scale based on an IR standard candle would be a
potent tool for determining distances of optically invisible dusty
galaxies.  Therefore, we have investigated whether the function by
\citet{jacoby89} also fits the histograms of IR magnitudes for LMC PNe.

\citet{henize63} first constructed a PNLF adopting a constant-mass,
uniformly expanding model for a PN whose central star did not evolve. They
also advanced the concept that PNe in different galaxies attain the same
maximum brightness.  The LF is an integrated snapshot of an evolving
ensemble of central stars of different mass and their associated evolving
nebulae.  \citet{ciard89} modified this expanding and slowly dimming model
to explain the sharp turn-down observed at the bright end of M31's PNLF.  
They included an exponential truncation to represent a cut-off in the
upper mass limit of PN central stars, and stellar evolution that is
strongly mass-dependent.  The same function still suffices to represent
PNLFs in the many galaxies for which a LF for the 5007\AA\, [\ion{O}{3}]
line (or an H$\alpha$ or H$\beta$ LF) has been built 
\citep[e.g.,][]{frew06}.  Optical PNLF histograms are fit by the function
N(M)=e$^{0.307M}$(1-e$^{3(M^{*}-M)})$ where M is the equivalent
[\ion{O}{3}] magnitude and M$^*$ the absolute 5007\AA\, magnitude of the
brightest PN that can exist in any galaxy.  M$^*$=$-$4.47$\pm$0.05 and
this is the [\ion{O}{3}] standard candle.  In fact M$^*$ depends slightly
on metallicity.  Rigorous testing has shown it to be fainter in metal-poor
galaxies but constant for galaxies whose [O/H] abundance is above that of
the LMC \citep{ciard05}.

\citet{ciard99} argued from the initial-mass-final mass-relation that
high-mass central stars must have massive circumstellar envelopes from
which dust cannot escape during the rapid changes in stellar conditions.  
Thus a PN around a massive central star will always suffer strong internal
extinction, diminishing its observed line flux, perhaps even preventing
its optical detection. \citet{ciard06} commented that although no PN is
observed above the [\ion{O}{3}] bright cut-off, PNe are certainly known
with true fluxes about twice that corresponding to M$^*$.  If the optical
cut-off were due to the combination of massive progenitors suffering
internal nebular extinction, then one would expect to observe IR LFs with
PNe brighter than any analogous IR cut-off because our LFs correspond to
wavelengths at which extinction is far smaller than at 5007\AA.  This is a
prediction that we can test.

PNe appear to emit in the IR by any or all of several possible mechanisms.
It is not obvious that they should mimic the behavior modeled for the
evolution of [\ion{O}{3}] and H$\alpha$ emission.  However, the fluorescence of
PAH bands, fine structure line radiation, and optically-thin thermal dust
emission are processes linked to stellar luminosity, particularly UV flux.  
Therefore, MIR luminosity in a PN may well track the variations in stellar
UV luminosity.  All PNe undergo an initial expansion and their central
stars are of low or intermediate mass, implying a declining mass spectrum
in their progenitor population.  Only for edge-on disk geometries would
the MIR emission of a PN also depend on the viewing angle and these are
associated with the roughly 10\% of all PNe that are Type\,1 N-rich
bipolar PNe and have the highest mass central stars. Consequently, an
initial exploration of the LMC infrared PNLFs is worthwhile.

\begin{figure*}
\includegraphics[angle=0, scale=0.38]{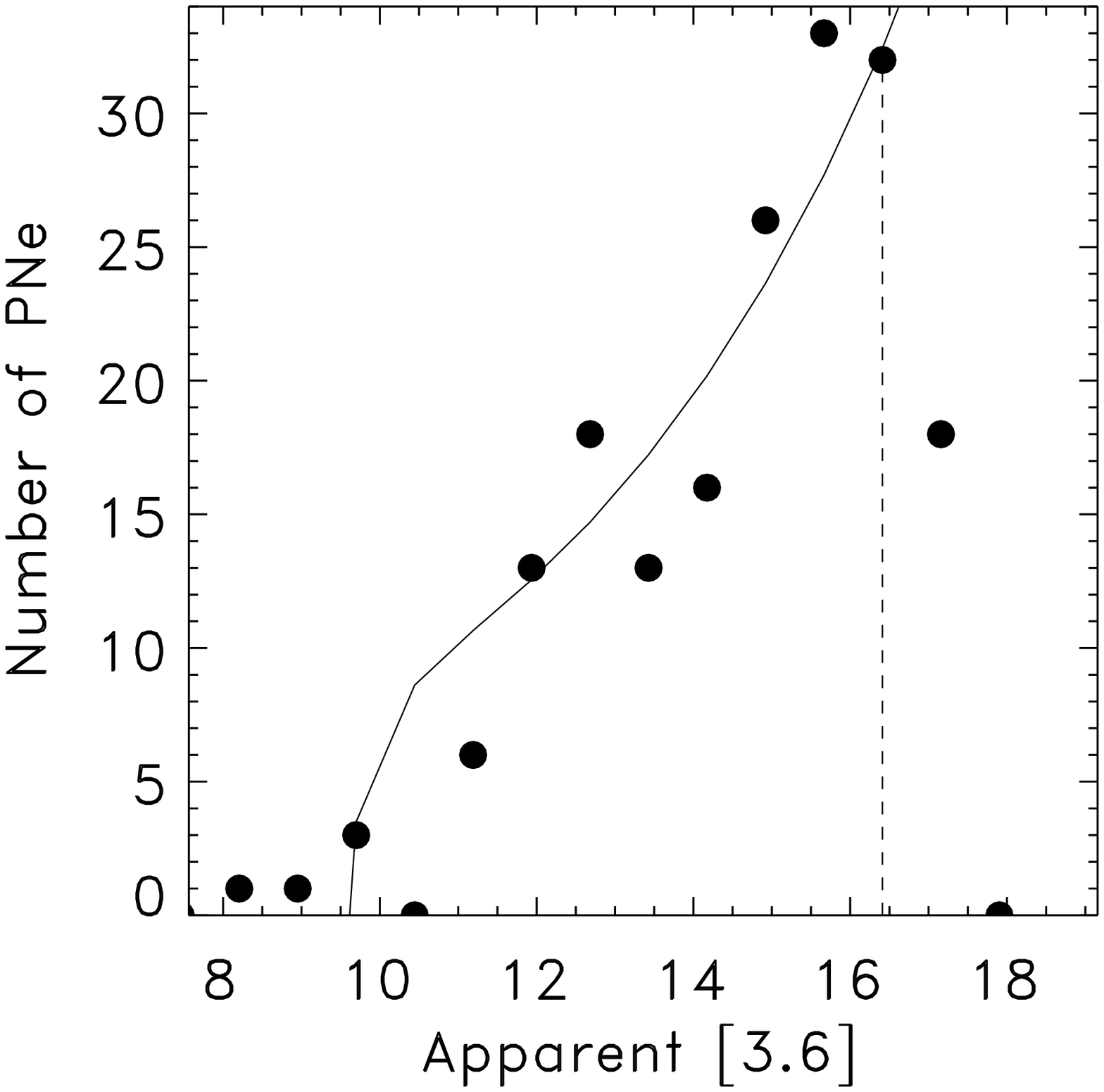}
\includegraphics[angle=0, scale=0.38]{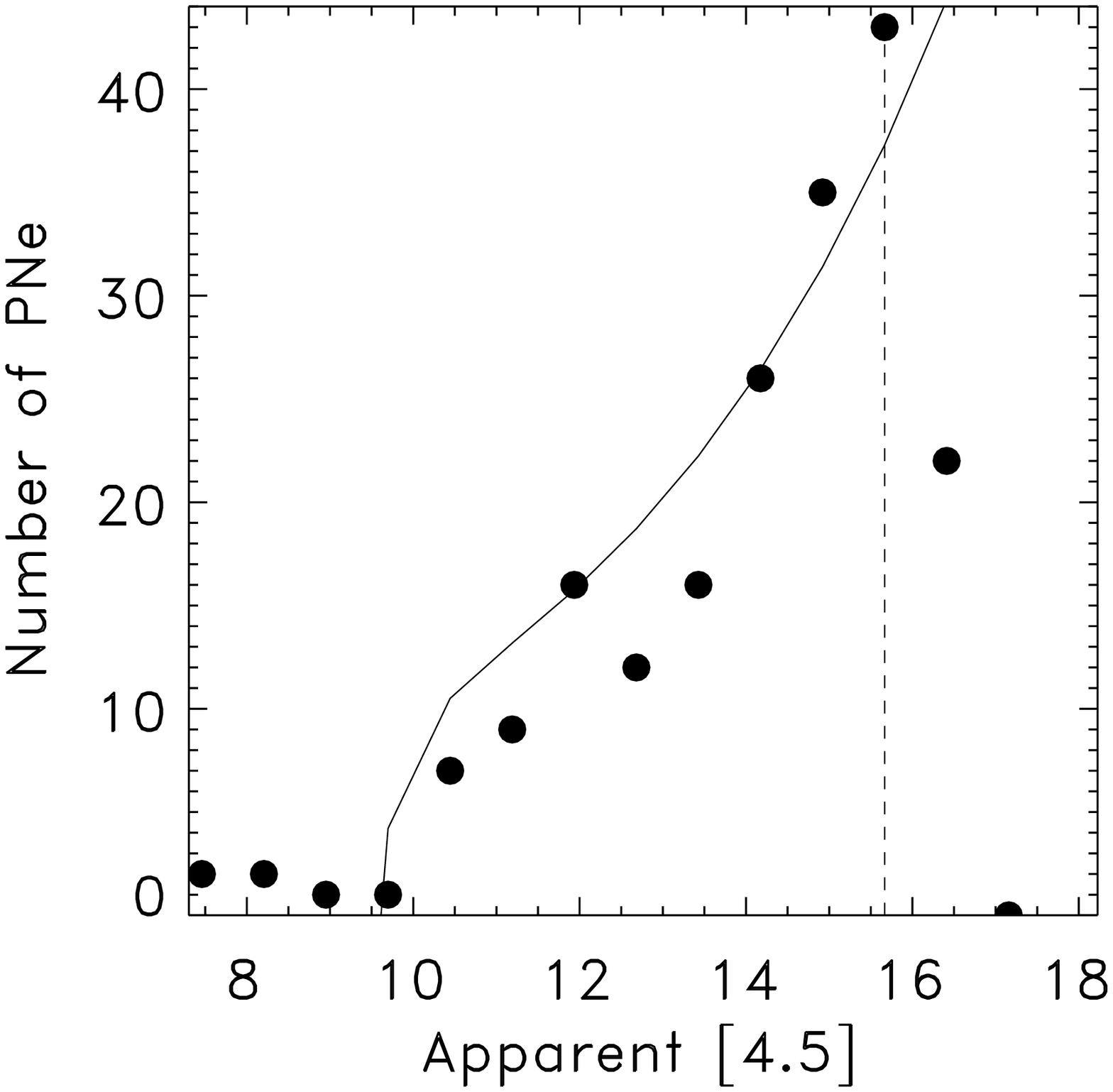}
\includegraphics[angle=0, scale=0.38]{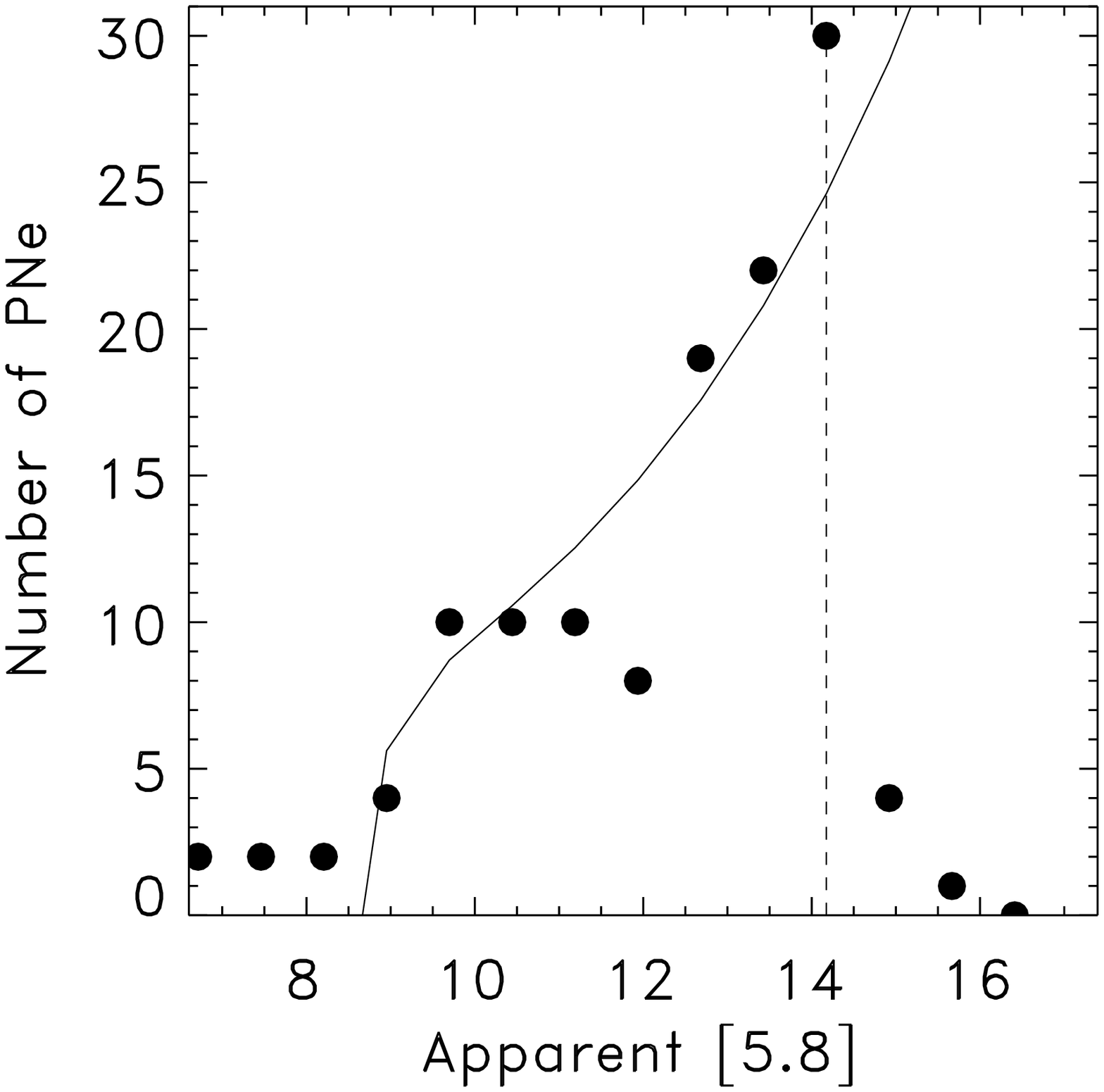}
\includegraphics[angle=0, scale=0.38]{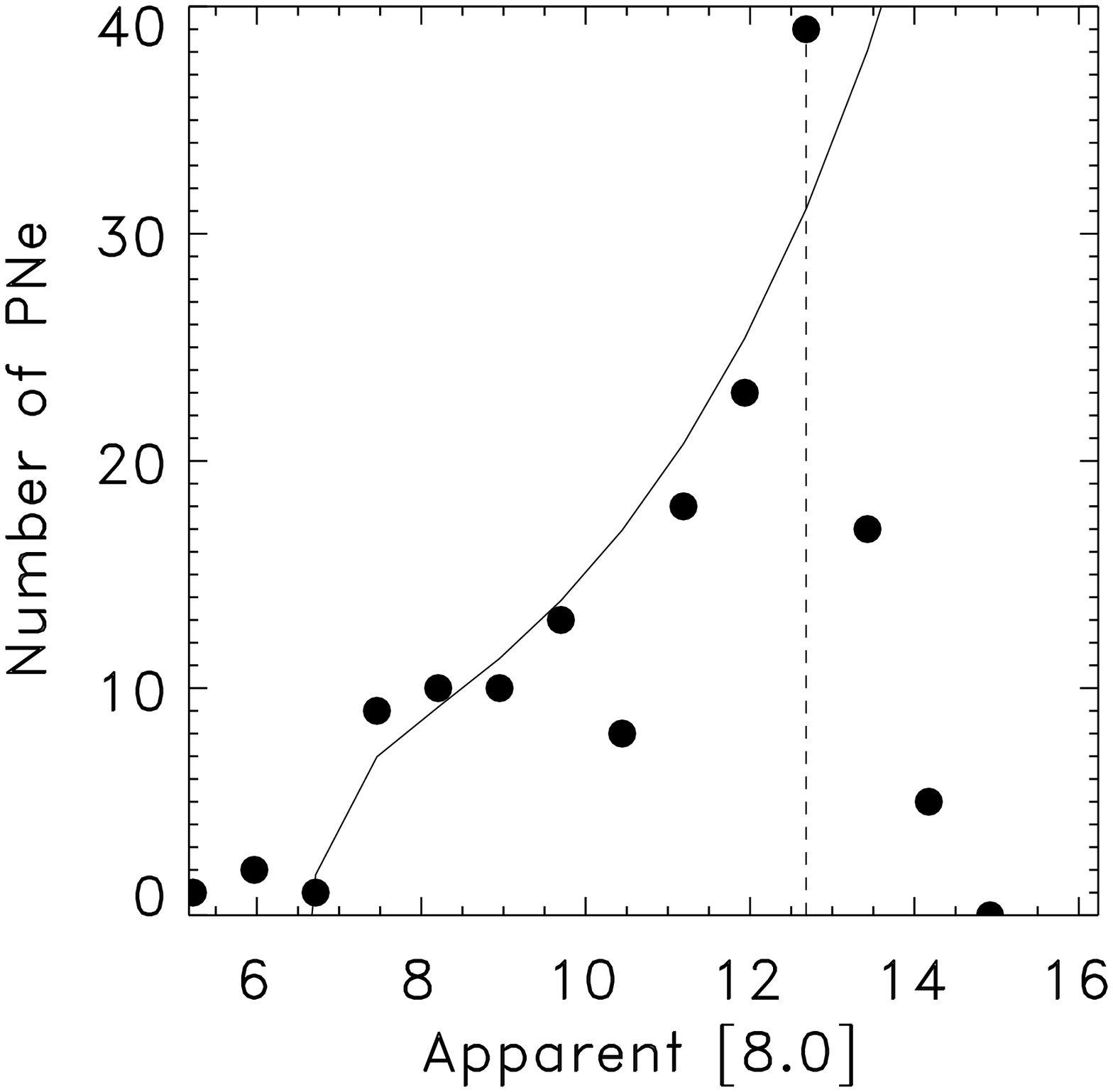}
\includegraphics[angle=0, scale=0.38]{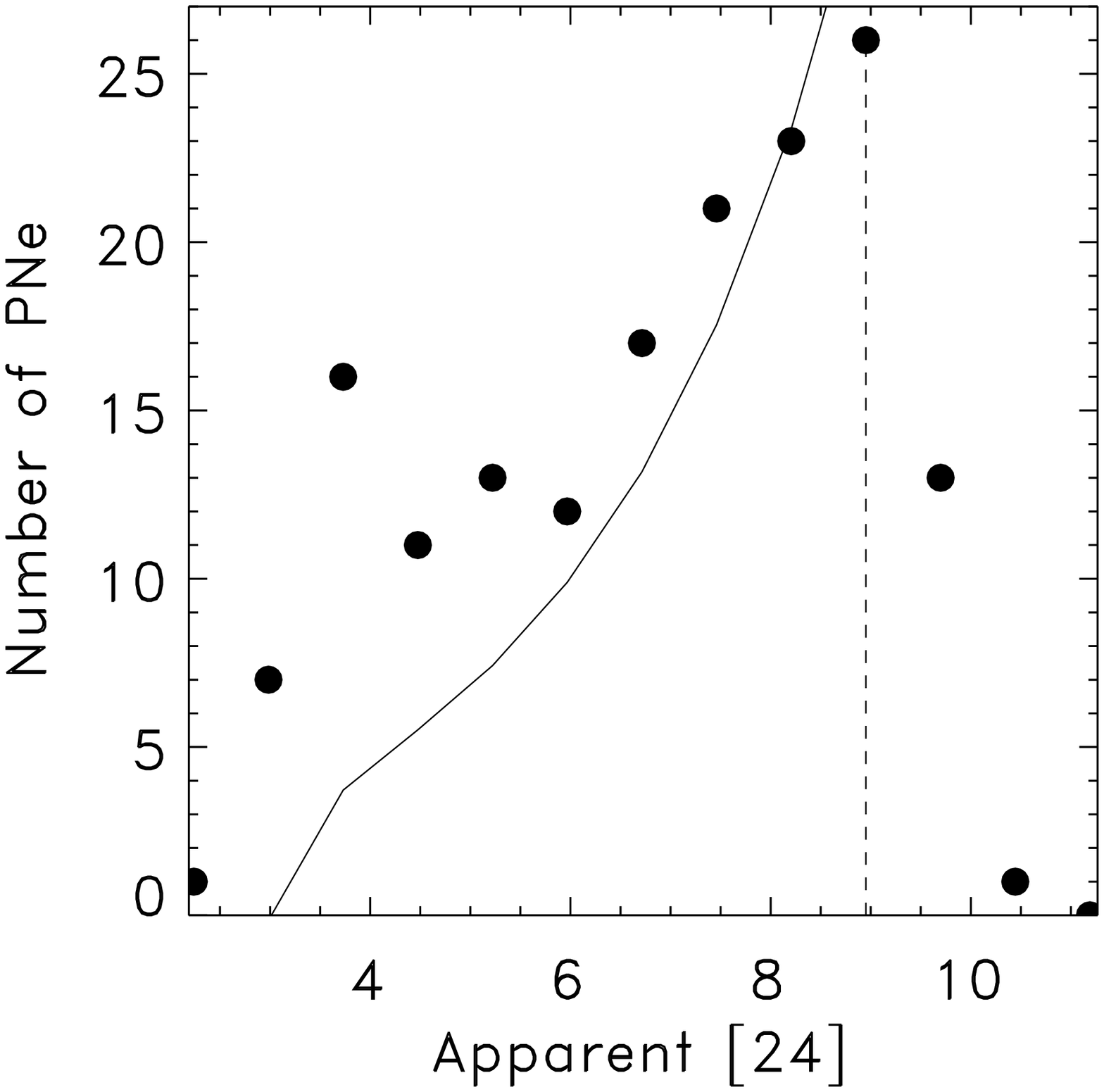}

\caption{The planetary nebulosity luminosity functions for each of the
IRAC channels and the MIPS 24 $\mu$m band. \label{clfs} The vertical dashed
line is the estimated completeness magnitude in each of the bands.}

\end{figure*}

Our magnitude samples are too sparse in the near-infrared and at 70 $\mu$m
but we present LFs in all IRAC bands and at 24 $\mu$m in Figure
\ref{clfs}.  We found that 0.5 magnitude bins are too small to remove the
noise but 0.75 magnitude appears to average out inappropriately high
frequency structure while preserving essential features. Each of the IRAC
LFs was fitted independently to the identical truncated exponential form
used by Jacoby and colleagues \citep[equation 2 of][]{ciard89}.  The
function was scaled by least squares minimization to best match the
observed counts in the domain from the obvious cut-off to the estimated
limit of completeness (the vertical dashed lines in these figures).  Each
IRAC LF resembles the optical LFs structurally in that we see cut-offs
near magnitudes of 9.5, 9.5, 8.5 and 6.6 in IRAC bands 1$-$4,
respectively.  We also note possible ``Jacoby'' dips near [3.6]=14.5,
[4.5]=13.5, [5.8]=12.0 and [8.0]=10.5, where the counts fall about a
factor of 2 below the exponential.  Any unpopulated regions in a PNLF are
attributed to populations in which central star evolution proceeds very
rapidly \citep{jacoby02}.  The IRAC PNLFs appear promising, with IR
analogs of the bright end cut-offs observed in optical emission lines.  
The quantitative connections between these IR magnitudes and the [\ion{O}{3}]
M$^*$ will be the subject of a future detailed study.

Of particular importance are those objects brighter than the cut-offs.
They may not all be PNe.  Spuriously matched field stars, symbiotic
objects, and \ion{H}{2} regions are the typical contaminants of PNLFs in
several wavelength regimes.  An inventory of all PNe that are brighter
than the cut-off at any IRAC wavelength shows only 6 different objects.  
These are LMC numbers 123, 125, 11, 63, 35, and 231, ordered from most to least
frequently encountered brighter than M$^*$.  Information on the
characteristics of these PNe from \citet{leisy97} and \citet{reid06} is
somewhat sparse but indicates that all are in the LMC (from radial
velocities), and that LMC 63 is regarded only as a ``possible PN'' by Reid
\& Parker because of its very-low excitation spectrum.  LMC11 (SMP11) is
more likely to be a post-AGB object or a proto-PN rather than a typical PN.
At present we cannot eliminate any of the other 4 objects as being non-PNe or  
transitional objects.  If any of these outlying PNe in our LFs survive
further scrutiny then this would validate the suggestion of heavy internal 
extinction in PNe with high-mass central stars.

We can offer no clear interpretation for the 24 $\mu$m LF.  Although it
looks superficially to mimic the universal optical form, our best-fitting
truncated exponential falls well below what appears to be a second peak in
the LF near [24]=3.5.  If we regard this as due to a LF dip near [24]=6
then it would argue for a far higher exponential contribution than the
observed counts support.  It is also possible that the secondary 24 $\mu$m
peak represents all the optically extinguished PNe beyond the [\ion{O}{3}]
cut-off that are now observable because of either the negligible
extinction at this long wavelength or the re-emission in the MIR of
internally absorbed optical luminosity.

\section{Conclusions}

We have presented images and photometry of the \citet{leisy97} sample of
PNe in the LMC.  Of the 233 known PNe in the survey field, 185 objects
were detected in at least two of the IRAC bands, and 161 detected in the
MIPS 24 $\mu$m images.  Color-color and color-magnitude diagrams were
presented using several combinations of IRAC, MIPS, and 2MASS magnitudes.  
The location of an individual PN in the color-color diagrams was seen to
depend on the relative contributions of the spectral components, resulting
in a wide range of colors for the objects in the sample.  A comparison to
a sample of Galactic PNe shows that they do not substantially differ in
their position in color-color space.  The location of PNe in the various
infrared color-color and color-magnitude diagrams are in general well
separated from normal stars, but overlap significantly with extragalactic
sources and potential YSOs.  Any ambiguity between PNe and YSOs or
galaxies can be readily resolved by the unique optical characteristics of
PNe and their environs.  Therefore, an IR color-based search for new PNe
in the LMC would be viable in combination with deep optical imaging and
spectroscopy. The latter remains the prerequisite to confirm a candidate
as a PN.

We have offered an exploration of the potential value of IR PNLFs in the
LMC. IRAC LFs appear to follow the same functional form as the
well-established [\ion{O}{3}] LFs although there are several PNe with
observed IR magnitudes brighter than the cut-offs in these LFs.  If these
objects can be demonstrated to be true PNe and not very-low excitation
variants nor symbiotic stars then their existence may confirm the
long-standing suggestion that PNe with massive central stars suffer heavy
internal extinction.  This extinction would eliminate optical outliers
beyond the cut-off magnitude but would affect IR LF counts minimally so
that all such outliers could be observed.

\acknowledgments

This work is based in part on observations made with the Spitzer Space
Telescope, which is operated by the Jet Propulsion Laboratory, California
Institute of Technology under NASA contract 1407. Support for this work
was provided by NASA through Contract Number 1256790 issued by
JPL/Caltech. Support for the IRAC instrument was provided by NASA through
Contract Number 960541 issued by JPL. This work is based in part on the
IRAC post-BCD processing software ``IRACproc'' developed by Mike
Schuster, Massimo Marengo and Brian Patten at the Smithsonian
Astrophysical Observatory.  MC thanks NASA for supporting his
participation in SAGE through JPL contract number 1275471 with UC
Berkeley. This work made use of the Two Micron All Sky Survey (2MASS)
database, which is a joint project of the University of Massachusetts and
the Infrared Processing and Analysis Center/California Institute of
Technology, funded by the National Aeronautics and Space Administration
and the National Science Foundation. This research has made use of the VizieR catalogue access tool and the SIMBAD database, CDS, Strasbourg, France.

\clearpage
\LongTables
\begin{landscape}
\begin{deluxetable}{lccccccccccc}
\tabletypesize{\scriptsize}
\tablecaption{SAGE LMC Planetary Nebulae Photometry \label{tbl-2}}
\tablewidth{0pt}
\tablehead{\colhead{LMC} & \colhead{Dist.\tablenotemark{a}} & \colhead{R.A.} & \colhead{Dec.} & 
\colhead{[3.6] (err)} & \colhead{[4.5] (err)} & \colhead{[5.8] (err)} & \colhead{[8.0] (err)} & 
\colhead{[24] (err)} & \colhead{[70] (err)} & \colhead{Other names} & \colhead{Code\tablenotemark{b}}   \\ 
\colhead{} & \colhead{(arcsec)} & \colhead{(J2000)} & \colhead{(J2000)} & \colhead{} & \colhead{} & 
\colhead{} & \colhead{} & \colhead{} & \colhead{} }
\startdata
1 & 0.18 & 4:38:34.8 & -70:36:43.3 & 12.34 (0.01) & 11.53 (0.01) & 10.16 (0.01) & 8.30 (0.00) & 4.03 (0.01) & 3.28 (0.11) & SMP1 & AAAAA   \\
2 & 0.74 & 4:40:56.7 & -67:48:01.6 & 13.65 (0.02) & 12.73 (0.02) & 12.05 (0.02) & 10.55 (0.01) & 7.42 (0.04) & \nodata & SMP2 & AAAAA   \\
5 & 1.2 & 4:48:08.7 & -67:26:06.3 & 13.21 (0.02) & 12.38 (0.01) & 11.61 (0.01) & 10.32 (0.01) & 6.68 (0.03) & \nodata & SMP5 &  AAAAA \\
7 & 0.75 & 4:48:29.7 & -69:08:12.6 & 14.85 (0.04) & 14.15 (0.03) & 13.50 (0.03) & 12.27 (0.03) & 5.96 (0.02) & \nodata & SMP7 & AAAAR \\
8 & 0.56 & 4:50:13.2 & -69:33:56.6 & 12.21 (0.01) & 11.15 (0.01) & 9.65 (0.01) & 8.12 (0.00) & 3.27 (0.01) & \nodata & SMP8 & AAAAA   \\
9 & 0.7 & 4:50:24.7 & -68:13:16.0 & 14.80 (0.03) & 14.61 (0.04) & 12.79 (0.02) & 11.09 (0.01) & 6.53 (0.04) & \nodata & SMP9 & AAAAA   \\
10 & 0.55 & 4:51:09.0 & -68:49:05.4 & 16.03 (0.07) & 15.16 (0.05) & \nodata & 13.69 (0.07) & 8.53 (0.05) & \nodata & SMP10 & AAAAA \\
11 & 0.62 & 4:51:37.9 & -67:05:16.3 & 10.35 (0.00) & 8.78 (0.00) & 7.35 (0.00) & 5.92 (0.00) & 2.94 (0.01) & 2.22 (0.09) & SMP11 & AAAAA   \\
12 & 0.28 & 4:52:01.5 & -68:39:16.9 & 16.45 (0.08) & 15.45 (0.06) & 14.68 (0.08) & 13.12 (0.05) & 7.37 (0.04) & \nodata & SMP12 & AAAAA \\
13 & 1.41 & 5:00:00.2 & -70:27:40.5 & 13.83 (0.02) & 13.04 (0.02) & 11.76 (0.01) & 9.76 (0.01) & 5.82 (0.03) & \nodata & SMP13 &  AAAAA \\
14 & 0.72 & 5:00:20.8 & -70:58:50.9 & 16.87 (0.09) & 16.01 (0.08) & \nodata & \nodata & 8.82 (0.06) & \nodata & SMP14 &  AAAAA \\
15 & 1.06 & 5:00:52.7 & -70:13:39.8 & 13.68 (0.02) & \nodata & \nodata & \nodata & 4.76 (0.02) & \nodata & SMP15 & AAAAA   \\
16 & 0.89 & 5:02:02.0 & -69:48:53.6 & 15.62 (0.05) & 14.80 (0.04) & 14.20 (0.05) & 12.41 (0.03) & 7.18 (0.04) & 3.47 (0.14) & SMP16 &  CCCAA \\
17 & 0.42 & 5:02:52.4 & -69:20:53.3 & 14.87 (0.04) & 13.88 (0.03) & 12.95 (0.02) & 11.36 (0.02) & 7.32 (0.04) & \nodata & SMP17 & AAAAA   \\
18 & 0.76 & 5:03:42.7 & -70:06:46.8 & 15.37 (0.05) & 14.78 (0.04) & 13.93 (0.05) & 12.38 (0.04) & 7.75 (0.05) & \nodata & SMP18 & BBAAA \\
19 & 0.58 & 5:03:41.3 & -70:13:53.0 & 14.09 (0.03) & 13.39 (0.02) & 11.96 (0.01) & 10.29 (0.01) & 5.39 (0.03) & \nodata & SMP19  & AAAAA \\
20 & 0.48 & 5:04:40.1 & -69:21:39.6 & 15.10 (0.04) & 14.75 (0.04) & 14.31 (0.06) & 13.11 (0.04) & 7.65 (0.04) & \nodata & SMP20 & BBAAA \\
21 & 1.81 & 5:04:52.1 & -68:39:08.6 & 14.45 (0.03) & 13.46 (0.02) & 12.19 (0.02) & 10.34 (0.01) & 4.76 (0.02) & 1.78 (0.06) & SMP21 & AAAAA   \\
22 & 0.89 & 5:05:50.7 & -69:02:31.1 & 15.64 (0.06) & 15.23 (0.06) & \nodata & \nodata & 9.13 (0.09) & \nodata & SMP22 & CCCCA \\
24 & 1.24 & 5:06:18.3 & -68:59:30.9 & 17.05 (0.13) & 16.25 (0.10) & \nodata & \nodata & 8.77 (0.05) & \nodata & SMP24 & NNNAA   \\
25 & 0.19 & 5:06:23.9 & -69:03:19.3 & 12.57 (0.01) & 12.23 (0.01) & 10.85 (0.01) & 9.02 (0.01) & 4.24 (0.01) & \nodata & SMP25 & AAAAA   \\
26 & 0.47 & 5:07:54.9 & -66:57:45.3 & 16.04 (0.06) & 15.33 (0.05) & 14.48 (0.06) & 12.42 (0.03) & 7.00 (0.04) & \nodata & SMP27 & EEEEA   \\
27 & 1.51 & 5:07:57.7 & -68:51:46.2 & 13.10 (0.02) & 11.97 (0.01) & 11.37 (0.01) & 10.39 (0.01) & 3.82 (0.02) & 1.21 (0.06) & SMP28 & AAAAA   \\
28 & 1.0 & 5:08:03.5 & -68:40:16.3 & 12.96 (0.02) & 11.92 (0.01) & 11.12 (0.01) & 9.30 (0.01) & 4.16 (0.01) & 1.92 (0.08) & SMP29 &  BBBAA \\
29 & 0.25 & 5:09:10.6 & -66:53:38.2 & 16.31 (0.07) & 15.20 (0.05) & 14.36 (0.06) & 13.24 (0.05) & 8.52 (0.05) & 3.32 (0.13) & SMP30 & AAAAA   \\
30 & 0.73 & 5:09:20.2 & -67:47:24.3 & 12.00 (0.01) & 11.05 (0.01) & 9.54 (0.00) & 7.63 (0.00) & 3.20 (0.01) & 2.30 (0.06) & SMP31 & AAAAA \\
31 & 0.74 & 5:09:37.2 & -70:49:08.1 & 14.65 (0.03) & 13.62 (0.02) & 12.82 (0.02) & 11.24 (0.02) & 5.65 (0.03) & \nodata & SMP32 & AAAAA \\
32 & 0.72 & 5:10:09.5 & -68:29:54.0 & 13.11 (0.02) & 12.74 (0.02) & 10.83 (0.01) & 9.06 (0.01) & 4.65 (0.02) & \nodata & SMP33 & AAAAA  \\
33 & 0.31 & 5:10:17.2 & -68:48:22.5 & 14.06 (0.03) & 13.43 (0.02) & 12.96 (0.03) & 11.05 (0.02) & 5.92 (0.02) & \nodata & SMP34 & AACCC \\
34 & 0.76 & 5:10:50.0 & -65:29:30.4 & 15.23 (0.04) & 14.27 (0.03) & 13.90 (0.04) & 12.61 (0.03) & 6.90 (0.04) & \nodata & SMP35 & AAAAA   \\
35 & 1.01 & 5:10:39.7 & -68:36:04.0 & 11.20 (0.01) & 10.63 (0.01) & 8.73 (0.00) & 6.89 (0.00) & 3.90 (0.02) & 1.97 (0.07) & SMP36 & CCBBB   \\
36 & 0.27 & 5:11:02.9 & -67:47:58.8 & 13.64 (0.02) & 13.06 (0.02) & 11.29 (0.01) & 9.49 (0.01) & 4.75 (0.01) & 3.15 (0.14) & SMP37  & AAAAA \\
37 & 1.19 & 5:11:23.8 & -70:01:56.5 & 12.18 (0.01) & 11.71 (0.01) & 9.79 (0.01) & 7.76 (0.00) & 4.33 (0.01) & 2.89 (0.09) & SMP38  & AAAAA \\
38 & 0.68 & 5:11:42.1 & -68:34:59.7 & 14.11 (0.03) & 13.58 (0.02) & 11.79 (0.01) & 9.89 (0.01) & 5.51 (0.03) & \nodata & SMP39 & AACCA \\
39 & 1.06 & 5:12:15.8 & -66:22:56.1 & 13.89 (0.02) & 13.28 (0.02) & 12.55 (0.02) & 11.25 (0.02) & 6.83 (0.03) & 3.26 (0.15) & SMP40 & AAAAA \\
40 & 0.43 & 5:13:27.3 & -70:33:34.7 & 15.73 (0.06) & 14.89 (0.04) & \nodata & 13.52 (0.06) & 7.30 (0.05) & \nodata & SMP41 & AAAAA \\
41 & 0.17 & 5:15:46.8 & -68:42:23.7 & 15.14 (0.04) & 14.37 (0.03) & 13.55 (0.04) & 11.73 (0.02) & 6.23 (0.03) & \nodata & SMP42  & AAAAA \\
43 & 0.91 & 5:18:29.9 & -67:16:55.8 & 15.32 (0.04) & 14.73 (0.04) & 13.90 (0.04) & 12.44 (0.03) & 7.58 (0.04) & \nodata & SMP44 & CCBBA   \\
44 & 0.19 & 5:19:20.7 & -66:58:07.5 & 15.00 (0.04) & 14.04 (0.03) & 13.66 (0.04) & 12.16 (0.03) & 6.59 (0.02) & \nodata & SMP45 & AAACA \\
45 & 1.1 & 5:19:29.7 & -68:51:07.9 & 14.98 (0.04) & 14.88 (0.05) & 12.83 (0.02) & 10.95 (0.01) & 6.97 (0.03) & \nodata & SMP46 & AAAAA \\
46 & 0.66 & 5:19:54.7 & -69:31:04.5 & 13.03 (0.02) & 12.10 (0.01) & 10.64 (0.01) & 8.81 (0.01) & 3.40 (0.01) & \nodata & SMP47 & BBCCA \\
47 & 0.92 & 5:20:09.7 & -69:53:38.9 & 11.73 (0.01) & 11.22 (0.01) & 10.39 (0.01) & 8.84 (0.01) & 4.40 (0.01) & \nodata & SMP48 & BBBBB   \\
48 & 0.23 & 5:20:09.4 & -70:25:38.1 & 16.39 (0.09) & 15.25 (0.06) & 15.83 (0.20) & 13.76 (0.07) & 8.02 (0.04) & \nodata & SMP49 & AAAAA   \\
49 & 1.0 & 5:20:51.7 & -67:05:42.5 & 13.32 (0.02) & 12.54 (0.01) & 11.98 (0.01) & 10.40 (0.01) & 5.42 (0.02) & 3.38 (0.17) & SMP50 & AAAAA \\
50 & 0.83 & 5:20:52.6 & -70:09:35.0 & 11.69 (0.01) & 10.74 (0.01) & 9.61 (0.01) & 7.54 (0.00) & 3.30 (0.01) & \nodata & SMP51 &  BBAAA \\
51 & 0.67 & 5:21:23.8 & -68:35:34.4 & 14.25 (0.03) & 13.24 (0.02) & 12.85 (0.02) & 10.93 (0.01) & 5.10 (0.02) & \nodata & SMP52  & AAAAA \\
52 & 0.49 & 5:21:32.9 & -67:00:04.0 & 14.72 (0.03) & 13.98 (0.03) & 13.61 (0.03) & 11.80 (0.02) & 5.00 (0.02) & 2.48 (0.08) & SMP53 & AAAAA   \\
53 & 0.19 & 5:21:42.9 & -68:39:24.8 & 16.08 (0.07) & 15.12 (0.05) & 14.27 (0.06) & 12.78 (0.05) & 7.85 (0.04) & \nodata & SMP54 & AAAAA   \\
54 & 0.2 & 5:22:41.0 & -71:19:06.7 & 12.48 (0.01) & 11.72 (0.01) & 10.52 (0.01) & 8.65 (0.00) & 4.15 (0.01) & 2.69 (0.07) & SMP55 & AAAEA   \\
55 & 0.46 & 5:23:31.2 & -69:04:04.4 & 13.20 (0.02) & 12.47 (0.01) & 11.77 (0.01) & 10.14 (0.01) & \nodata & \nodata & SMP56 & CCCCA   \\
56 & 0.7 & 5:23:48.6 & -69:12:21.6 & \nodata & 14.58 (0.04) & \nodata & \nodata & 8.15 (0.05) & \nodata & SMP57 & CCCCA \\
57 & 1.15 & 5:24:20.9 & -70:05:00.5 & 12.16 (0.01) & 11.02 (0.01) & 9.87 (0.01) & 8.05 (0.00) & 3.99 (0.01) & \nodata & SMP58 &  AAAAA \\
58 & 0.91 & 5:24:27.4 & -70:22:23.7 & 15.49 (0.05) & 14.60 (0.04) & 13.96 (0.04) & 12.71 (0.03) & 8.65 (0.05) & \nodata & SMP59 & AAEEA   \\
59 & 0.28 & 5:24:15.7 & -70:53:56.3 & 14.98 (0.04) & 14.32 (0.03) & 13.22 (0.03) & 11.20 (0.01) & 6.66 (0.02) & \nodata & SMP60 & CCCEA   \\
61 & 0.97 & 5:24:55.0 & -71:32:55.4 & 13.68 (0.02) & 12.60 (0.02) & 11.97 (0.02) & 9.83 (0.01) & 3.55 (0.01) & 2.05 (0.06) & SMP62  & AAAAA \\
62 & 1.09 & 5:25:26.2 & -68:55:53.8 & 13.67 (0.02) & 12.81 (0.02) & 12.13 (0.02) & 10.28 (0.01) & 5.11 (0.02) & \nodata & SMP63 & AAAAA \\
63 & 0.88 & 5:27:35.8 & -69:08:56.3 & 9.88 (0.00) & 8.73 (0.00) & 7.82 (0.00) & 6.60 (0.00) & \nodata & \nodata & SMP64 & CCCBB  \\
64 & 0.47 & 5:27:43.8 & -71:25:56.0 & 14.67 (0.03) & 14.00 (0.03) & 13.33 (0.04) & 11.55 (0.02) & 7.54 (0.07) & \nodata & SMP65 & AAAAA   \\
65 & 1.22 & 5:28:41.2 & -67:33:39.1 & 14.95 (0.04) & 14.20 (0.03) & 13.45 (0.03) & 11.55 (0.02) & 6.30 (0.03) & \nodata & SMP66 & BBAAA \\
66 & 0.67 & 5:29:15.7 & -67:32:46.7 & 14.74 (0.03) & 14.35 (0.03) & 12.88 (0.02) & 11.04 (0.01) & 5.71 (0.03) & \nodata & SMP67 & AAAAA \\
67 & 0.52 & 5:29:02.9 & -70:19:24.8 & 15.91 (0.08) & 14.30 (0.04) & 14.07 (0.05) & 12.67 (0.03) & 6.20 (0.03) & \nodata & SMP68 & AAAAA \\
68 & 0.61 & 5:29:23.2 & -67:13:21.9 & 16.40 (0.08) & 15.44 (0.06) & 14.62 (0.09) & 13.19 (0.06) & 8.00 (0.05) & \nodata & SMP69 & AAAAA \\
69 & 0.55 & 5:29:26.6 & -72:38:42.6 & 16.17 (0.07) & 15.28 (0.05) & 14.18 (0.05) & 13.11 (0.04) & 8.06 (0.05) & \nodata & SMP70 & AAAAA   \\
70 & 0.43 & 5:30:33.3 & -70:44:37.6 & 12.83 (0.01) & 12.41 (0.01) & 10.65 (0.01) & 8.84 (0.01) & 4.53 (0.01) & \nodata & SMP71 & AAAAA \\
71 & 0.94 & 5:30:45.8 & -70:50:15.8 & 13.42 (0.02) & 13.59 (0.02) & 13.40 (0.03) & 13.17 (0.04) & 7.91 (0.04) & \nodata & SMP72 & AAAAA   \\
72 & 0.92 & 5:31:22.0 & -70:40:44.9 & 12.35 (0.01) & 11.88 (0.01) & 10.23 (0.01) & 8.42 (0.00) & 4.22 (0.02) & \nodata & SMP73 & AAAAA \\
73 & 0.34 & 5:33:29.8 & -71:52:28.4 & 13.20 (0.02) & 12.53 (0.01) & 11.07 (0.01) & 9.10 (0.01) & 5.07 (0.02) & 3.26 (0.13) & SMP74 & AAAAA \\
74 & 0.17 & 5:33:47.0 & -68:36:44.2 & 11.81 (0.01) & 11.29 (0.01) & 9.35 (0.00) & 7.46 (0.00) & 4.03 (0.01) & \nodata & SMP75 & AAAAA \\
75 & 0.79 & 5:33:56.2 & -67:53:08.3 & 13.26 (0.02) & 12.45 (0.01) & 10.95 (0.01) & 9.04 (0.01) & 4.68 (0.01) & \nodata & SMP76 &  AAAAA \\
76 & 1.1 & 5:34:06.3 & -69:26:17.7 & 13.09 (0.02) & 12.37 (0.01) & 11.64 (0.01) & 9.96 (0.01) & 5.89 (0.03) & \nodata & SMP77 &  CCAAA \\
77 & 0.62 & 5:34:21.3 & -68:58:24.9 & 12.73 (0.01) & 11.99 (0.01) & 10.39 (0.01) & 8.36 (0.00) & 4.02 (0.02) & \nodata & SMP78 &  AAAAA \\
79 & 1.43 & 5:34:39.0 & -70:19:55.5 & 15.15 (0.04) & 14.71 (0.04) & \nodata & 13.86 (0.08) & 8.98 (0.07) & \nodata & SMP80 & CAAAA  \\
81 & 0.41 & 5:35:57.6 & -69:58:16.6 & 15.15 (0.04) & 14.36 (0.04) & 13.14 (0.03) & 11.40 (0.02) & 5.33 (0.02) & \nodata & SMP82 & CCCCA   \\
82 & 0.21 & 5:36:20.8 & -67:18:07.5 & 14.73 (0.03) & 13.71 (0.02) & 13.53 (0.03) & 11.80 (0.02) & 5.22 (0.02) & \nodata & SMP83 & AAAAA \\
83 & 0.43 & 5:36:53.0 & -71:53:38.0 & 14.50 (0.03) & 13.41 (0.02) & \nodata & 11.06 (0.01) & 6.15 (0.03) & \nodata & SMP84 & AAAAA \\
84 & 0.68 & 5:40:30.9 & -66:17:37.0 & 12.19 (0.01) & 11.41 (0.01) & 9.84 (0.01) & 8.01 (0.00) & 3.44 (0.02) & \nodata & SMP85 & AAAAA \\
85 & 0.54 & 5:41:22.1 & -68:07:44.2 & 16.56 (0.08) & 15.70 (0.07) & 15.03 (0.08) & 14.04 (0.07) & 9.03 (0.08) & \nodata & SMP86 &  AAAAA \\
86 & 0.84 & 5:41:08.0 & -72:42:07.8 & 15.33 (0.04) & 14.44 (0.03) & 13.33 (0.03) & 11.80 (0.02) & \nodata & 2.67 (0.10) & SMP87 & AAAEA   \\
87 & 0.56 & 5:42:33.3 & -70:29:24.0 & 13.33 (0.02) & 12.74 (0.02) & 12.46 (0.02) & 11.46 (0.02) & 7.47 (0.03) & \nodata & SMP88 & AAAAA \\
88 & 1.92 & 5:42:37.0 & -70:09:31.1 & 12.62 (0.01) & 12.26 (0.01) & 10.43 (0.01) & 8.80 (0.01) & 4.27 (0.01) & \nodata & SMP89 & AAAAA \\
89 & 0.9 & 5:44:34.8 & -70:21:40.0 & 13.35 (0.02) & 12.45 (0.01) & 11.62 (0.01) & 10.39 (0.01) & 7.08 (0.04) & \nodata & SMP90 & AAAAA \\
90 & 0.16 & 5:45:06.0 & -68:06:50.9 & 16.34 (0.07) & 15.47 (0.06) & \nodata & 13.18 (0.05) & 8.45 (0.06) & \nodata & SMP91 &  AAAAA \\
91 & 0.96 & 5:47:04.7 & -69:27:33.3 & 12.60 (0.01) & 12.10 (0.01) & 11.00 (0.01) & 9.31 (0.01) & 4.21 (0.01) & \nodata & SMP92 & BBBBB \\
92\tablenotemark{d} & 0.94 & 5:49:38.8 & -69:09:59.3 & 14.72 (0.03) & 13.72 (0.02) & 12.10 (0.02)& 10.90 (0.02) & 8.09 (0.03) & \nodata & SMP93 & EEEEE \\
94 & 0.58 & 6:01:45.3 & -67:56:06.4 & 15.48 (0.05) & 15.08 (0.05) & 13.40 (0.03) & 11.66 (0.02) & 7.42 (0.04) & \nodata & SMP95 & AAAAA \\
102 & 1.56 & 4:24:37.6 & -69:42:20.7 & 12.47 (0.01) & 11.67 (0.01) & 11.38 (0.01) & 10.75 (0.01) & \nodata & \nodata & Sa104  & AAAAN \\
103 & 0.69 & 5:02:32.9 & -69:26:15.2 & 16.30 (0.08) & 15.51 (0.06) & 14.19 (0.05) & 12.85 (0.04) & 9.09 (0.11) & \nodata & Sa105  & AAAAB \\
104 & 0.95 & 5:03:05.8 & -68:33:37.0 & 16.01 (0.06) & 15.86 (0.07) & 14.14 (0.05) & 12.33 (0.03) & 9.14 (0.08) & \nodata & Sa106  & CCCAA \\
105 & 1.13 & 5:06:43.7 & -69:15:37.8 & 15.53 (0.05) & 15.27 (0.06) & 14.50 (0.07) & 13.02 (0.05) & 8.43 (0.04) & \nodata & Sa107  & CCCAA \\
106 & 1.26 & 5:11:48.1 & -69:23:43.1 & 11.85 (0.01) & 11.12 (0.01) & 10.52 (0.01) & 9.37 (0.01) & 6.22 (0.04) & \nodata & Sa109  & AAAAA \\
107 & 1.05 & 5:12:16.7 & -68:29:10.1 & \nodata & 16.96 (0.15) & \nodata & \nodata & 10.20 (0.18) & \nodata & Sa110 & NNNNN  \\
108 & 0.36 & 5:24:56.7 & -69:15:31.2 & 15.69 (0.06) & 14.50 (0.04) & 13.69 (0.06) & 11.66 (0.04) & 6.51 (0.03) & \nodata & Sa117 & AAACC   \\
109 & 0.31 & 5:29:32.7 & -70:17:39.0 & \nodata & 16.17 (0.09) & \nodata & 13.24 (0.05) & 9.10 (0.07) & \nodata & Sa120 & NNAAA   \\
110 & 0.66 & 5:30:26.3 & -71:13:48.0 & 17.81 (0.17) & 17.15 (0.14) & \nodata & \nodata & \nodata & \nodata & Sa121 & NNNNN   \\
111 & 0.67 & 5:34:24.3 & -69:34:28.0 & \nodata & 15.65 (0.07) & 15.66 (0.21) & \nodata & 7.33 (0.04) & \nodata & Sa122 &  NNNNC \\
112 & 0.35 & 5:34:30.2 & -70:28:34.5 & 15.32 (0.05) & 14.70 (0.04) & 13.31 (0.03) & 11.46 (0.02) & 6.71 (0.03) & \nodata & Sa123  & AAAAA \\
113 & 0.68 & 5:40:44.6 & -67:18:07.7 & 16.51 (0.08) & 15.66 (0.06) & 14.55 (0.07) & 13.00 (0.04) & 8.94 (0.07) & \nodata & Sa124  & AAAAA \\
114 & 0.58 & 5:53:14.6 & -70:25:02.0 & 15.81 (0.06) & 15.67 (0.06) & 14.40 (0.06) & 12.70 (0.03) & 8.46 (0.06) & \nodata & Sa126 & AAAAA  \\
118 & 0.44 & 5:12:59.9 & -68:57:07.7 & 18.41 (0.37) & 17.38 (0.22) & \nodata & \nodata & 9.78 (0.12) & \nodata & J10 & NNNNA \\
120 & 0.94 & 5:15:08.7 & -69:21:00.9 & \nodata & \nodata & \nodata & 13.37 (0.07) & 9.95 (0.16) & \nodata & J14 & NNNCC \\
123 & 1.48 & 5:17:00.7 & -69:19:29.1 & 8.96 (0.00) & 8.05 (0.00) & 7.43 (0.00) & 6.53 (0.00) & 4.39 (0.01) & \nodata & J17 & AAAAA   \\
124 & 0.46 & 5:17:23.9 & -69:39:13.2 & 15.83 (0.07) & \nodata & \nodata & \nodata & \nodata & \nodata & J18 & NNNNN   \\
125 & 0.38 & 5:17:58.9 & -69:39:23.0 & 8.31 (0.00) & 8.19 (0.00) & 7.83 (0.00) & 7.59 (0.00) & 5.59 (0.02) & \nodata & J19 & CCCCC   \\
126 & 1.1 & 5:18:45.8 & -69:10:11.5 & 15.73 (0.07) & \nodata & \nodata & \nodata & \nodata & \nodata & Sa112 & NNNNN   \\
127 & 0.22 & 5:18:55.7 & -69:33:02.1 & 13.60 (0.02) & 13.81 (0.03) & 13.81 (0.05) & 13.88 (0.16) & 9.84 (0.13) & \nodata & J21 & AAAAN   \\
128 & 0.85 & 5:19:07.0 & -69:41:54.0 & 12.79 (0.01) & 12.82 (0.02) & 12.71 (0.02) & 12.22 (0.03) & 8.35 (0.06) & \nodata & J22 & CCCCC   \\
129 & 0.26 & 5:19:15.1 & -69:34:52.6 & 15.59 (0.08) & 15.68 (0.09) & \nodata & \nodata & \nodata & \nodata & J23 & AACCC   \\
132 & 0.17 & 5:20:56.9 & -70:05:10.5 & 13.56 (0.02) & 13.65 (0.03) & 13.50 (0.04) & \nodata & \nodata & \nodata & Sa114 & mis-ID? \\
133 & 1.9 & 5:21:07.6 & -69:44:28.1 & \nodata & 16.01 (0.13) & 14.65 (0.09) & 12.49 (0.03) & 9.27 (0.09) & \nodata & J32 & NNAAC   \\
134 & 1.48 & 5:21:17.6 & -69:43:01.0 & \nodata & \nodata & \nodata & 12.66 (0.07) & 7.54 (0.04) & \nodata & Sa115 & NNNCB   \\
135 & 1.44 & 5:24:36.6 & -69:05:51.1 & 15.42 (0.05) & 14.87 (0.04) & 14.01 (0.05) & 12.88 (0.05) & 8.44 (0.05) & \nodata & Sa116 & AAAAA   \\
136 & 1.66 & 5:26:09.5 & -69:00:58.5 & 15.22 (0.04) & 15.06 (0.05) & \nodata & \nodata & 8.66 (0.06) & \nodata & Sa118 & BBAAA \\
137 & 0.93 & 4:47:22.8 & -67:41:18.8 & \nodata & 16.57 (0.10) & \nodata & \nodata & \nodata & \nodata & MG01 & NNNNN   \\
138 & 0.89 & 4:48:09.3 & -68:33:39.0 & 16.52 (0.08) & 15.74 (0.07) & 14.42 (0.06) & 13.07 (0.04) & \nodata & \nodata & MG02 &  AAAAA \\
139 & 0.83 & 4:50:56.7 & -66:19:52.5 & 16.99 (0.10) & 16.71 (0.11) & \nodata & 13.13 (0.04) & 10.26 (0.17) & \nodata & MG03 & NNAAN   \\
140 & 0.72 & 4:52:45.2 & -70:17:49.1 & \nodata & 16.40 (0.10) & \nodata & \nodata & \nodata & \nodata & MG04 & NNAAN   \\
141 & 0.69 & 4:53:29.7 & -68:22:52.3 & 17.87 (0.20) & 16.79 (0.13) & \nodata & \nodata & 9.24 (0.08) & \nodata & MG05 & NNNAA \\
144 & 1.28 & 4:55:39.8 & -68:34:20.3 & 17.28 (0.14) & 16.18 (0.09) & \nodata & 13.99 (0.17) & 9.50 (0.10) & \nodata & MG08 & NNAAA   \\
145 & 1.56 & 4:56:21.1 & -67:24:22.5 & 16.81 (0.10) & 16.05 (0.08) & 14.63 (0.07) & \nodata & \nodata & \nodata & MG09 & NNAAN   \\
146 & 1.11 & 4:56:37.8 & -67:40:54.6 & 16.46 (0.08) & 16.28 (0.09) & 14.18 (0.05) & 12.30 (0.03) & 9.05 (0.07) & \nodata & MG10 & AAAAA \\
147 & 0.66 & 4:59:18.3 & -67:27:03.9 & 17.15 (0.12) & 16.41 (0.10) & \nodata & \nodata & 8.40 (0.05) & \nodata & MG11 & NNNAA   \\
148 & 0.92 & 5:01:40.3 & -66:46:45.6 & \nodata & 17.28 (0.16) & \nodata & \nodata & \nodata & \nodata & MG12 & NNNNN   \\
149 & 0.57 & 5:03:03.3 & -65:23:02.3 & 18.22 (0.20) & 16.93 (0.12) & \nodata & 14.61 (0.11) & 10.17 (0.16) & \nodata & MG13 & NNNNN   \\
150 & 0.52 & 5:04:27.7 & -68:58:11.4 & 15.35 (0.05) & 14.67 (0.04) & 13.75 (0.05) & 11.65 (0.03) & 7.31 (0.03) & \nodata & MG14  & CCCCC \\
151 & 1.31 & 5:05:35.7 & -68:01:40.1 & 16.71 (0.09) & 16.25 (0.09) & \nodata & \nodata & \nodata & \nodata & MG15 & NNNEN   \\
152 & 0.06 & 5:06:05.2 & -64:48:49.2 & 16.80 (0.09) & 15.97 (0.07) & \nodata & \nodata & \nodata & \nodata & MG16 & AAAAA \\
153 & 0.39 & 5:06:21.2 & -64:37:03.5 & 17.19 (0.11) & 16.67 (0.11) & 14.81 (0.08) & 13.28 (0.04) & 8.33 (0.06) & \nodata & MG17 & AAAAA   \\
155 & 0.97 & 5:08:32.0 & -68:09:44.3 & 17.19 (0.12) & 16.60 (0.11) & \nodata & 13.42 (0.06) & \nodata & \nodata & MG19 & CCCCN   \\
156 & 0.5 & 5:10:40.2 & -68:10:23.5 & 17.26 (0.12) & 16.70 (0.11) & \nodata & \nodata & \nodata & \nodata & MG20 & NNNNN   \\
157 & 0.84 & 5:11:38.0 & -65:42:42.1 & 16.88 (0.09) & 15.77 (0.07) & \nodata & \nodata & 7.78 (0.04) & \nodata & MG21 & NAAAA   \\
158 & 0.98 & 5:11:53.5 & -65:32:27.0 & 18.16 (0.20) & 17.48 (0.18) & \nodata & \nodata & 9.56 (0.11) & \nodata & MG22 & NNNNA   \\
159 & 0.85 & 5:11:47.4 & -68:16:09.5 & 16.72 (0.09) & 16.25 (0.09) & \nodata & 13.54 (0.07) & 8.47 (0.05) & \nodata & MG23 & CCCCA   \\
160 & 0.06 & 5:13:01.0 & -65:15:34.1 & 18.02 (0.18) & \nodata & \nodata & 15.49 (0.25) & \nodata & \nodata & MG24 & NNNNN   \\
161 & 0.79 & 5:12:17.8 & -71:54:49.5 & 17.46 (0.13) & 16.83 (0.12) & \nodata & \nodata & \nodata & \nodata & MG25 & AANNN   \\
162 & 0.46 & 5:13:28.0 & -66:17:28.4 & 13.28 (0.02) & 12.51 (0.01) & 11.23 (0.01) & 9.18 (0.01) & 5.09 (0.03) & \nodata & MG26 & AAAAA \\
163 & 0.29 & 5:13:34.2 & -65:35:13.2 & 17.83 (0.16) & \nodata & \nodata & \nodata & 9.91 (0.13) & \nodata & MG27 & NNNNN   \\
164 & 0.48 & 5:13:09.7 & -69:31:18.0 & \nodata & \nodata & \nodata & 14.62 (0.19) & 9.91 (0.16) & \nodata & MG28 & CCCCC \\
165 & 0.88 & 5:13:42.4 & -68:15:16.4 & 16.34 (0.08) & 15.58 (0.07) & \nodata & \nodata & 7.68 (0.04) & \nodata & MG29 & CAAAC   \\
166 & 0.32 & 5:14:14.7 & -70:50:31.5 & 16.20 (0.07) & 15.27 (0.06) & 14.70 (0.08) & 12.77 (0.03) & 7.21 (0.03) & \nodata & MG30 & CCAAA   \\
167 & 0.75 & 5:16:29.3 & -68:18:11.3 & 16.67 (0.09) & 16.20 (0.09) & 14.62 (0.08) & 12.83 (0.04) & 8.93 (0.06) & \nodata & MG31 & CCCAA   \\
168 & 0.4 & 5:19:04.0 & -64:59:18.6 & 16.19 (0.07) & 15.55 (0.06) & \nodata & 13.57 (0.06) & \nodata & \nodata & MG32 & AAAAN   \\
169 & 1.11 & 5:19:13.7 & -66:09:31.3 & 18.00 (0.18) & 17.18 (0.14) & 16.65 (0.37) & 14.64 (0.12) & 10.52 (0.24) & \nodata & MG33 & NNNEA \\
170 & 0.61 & 5:19:28.9 & -67:14:26.5 & 16.22 (0.07) & 15.47 (0.06) & 13.97 (0.05) & 12.08 (0.02) & 8.07 (0.05) & \nodata & MG34 & AAAAA   \\
171 & 0.47 & 5:19:33.4 & -66:55:37.2 & 11.26 (0.01) & 10.38 (0.01) & 9.71 (0.01) & 8.68 (0.00) & 6.04 (0.03) & \nodata & MG35 & AAAAA \\
172 & 0.9 & 5:20:29.7 & -64:53:14.0 & 16.51 (0.08) & 15.53 (0.06) & \nodata & 13.16 (0.04) & \nodata & \nodata & MG36 & EEEEA \\
173 & 0.46 & 5:21:46.8 & -65:22:26.7 & 15.77 (0.06) & 15.18 (0.05) & 14.43 (0.06) & 12.68 (0.03) & 8.33 (0.05) & \nodata & MG37 & AAAAA   \\
174 & 0.65 & 5:22:03.0 & -64:25:16.0 & 17.49 (0.14) & \nodata & \nodata & \nodata & \nodata & \nodata & MG38 & NNNAA   \\
175 & 0.63 & 5:22:12.9 & -69:43:29.1 & 12.42 (0.01) & 11.15 (0.01) & 9.92 (0.01) & 8.30 (0.01) & 3.50 (0.01) & \nodata & MG39 & AACCC \\
176 & 0.62 & 5:22:35.3 & -68:24:25.3 & 15.72 (0.06) & 14.90 (0.04) & 14.19 (0.06) & 12.17 (0.03) & 7.03 (0.03) & \nodata & MG40 & CCCCA \\
177 & 0.61 & 5:23:37.7 & -65:09:54.3 & 16.41 (0.07) & 16.21 (0.08) & 13.97 (0.04) & 12.16 (0.02) & 9.16 (0.09) & \nodata & MG41 & AAAAA \\
179 & 0.1 & 5:24:34.2 & -71:13:39.5 & \nodata & 16.46 (0.10) & \nodata & 14.13 (0.09) & 8.41 (0.06) & \nodata & MG43 & NNNAA   \\
182 & 0.9 & 5:26:20.5 & -65:21:45.9 & 14.33 (0.03) & 13.96 (0.03) & 13.73 (0.04) & 13.13 (0.04) & \nodata & \nodata & MG46 & AAAAA   \\
183 & 1.8 & 5:26:45.3 & -64:38:00.8 & 14.87 (0.04) & 13.87 (0.03) & 12.73 (0.02) & 10.31 (0.01) & 8.98 (0.07) & \nodata & MG47 & AAAAA \\
184 & 0.76 & 5:26:59.8 & -66:07:05.8 & 17.27 (0.11) & 16.50 (0.10) & \nodata & 13.50 (0.05) & 9.01 (0.08) & \nodata & MG48 & AANAA   \\
185 & 0.08 & 5:27:32.1 & -69:32:18.8 & 15.75 (0.06) & 14.92 (0.05) & 14.43 (0.06) & 12.61 (0.03) & 7.11 (0.03) & \nodata & MG49 & CCCAA \\
186 & 0.21 & 5:29:08.6 & -66:42:56.3 & 17.80 (0.16) & 16.12 (0.08) & \nodata & \nodata & 7.56 (0.04) & \nodata & MG50 & NANAA \\
187 & 0.22 & 5:28:34.4 & -70:33:01.6 & 15.23 (0.04) & 14.93 (0.05) & \nodata & 12.62 (0.03) & 9.15 (0.10) & \nodata & MG51 & CCAAA   \\
188 & 0.54 & 5:29:08.8 & -69:45:28.0 & 16.15 (0.08) & 15.78 (0.08) & 14.31 (0.06) & 12.54 (0.03) & 8.24 (0.05) & \nodata & MG52 & CCAAA   \\
189 & 0.78 & 5:29:35.7 & -69:46:05.2 & \nodata & 15.30 (0.06) & \nodata & \nodata & 9.35 (0.10) & \nodata & MG53 & CCCCA   \\
190 & 0.69 & 5:29:51.4 & -68:50:05.6 & 16.03 (0.06) & 15.57 (0.06) & \nodata & 12.88 (0.04) & \nodata & \nodata & MG54 & CCCAN   \\
191 & 1.1 & 5:31:09.1 & -71:36:40.2 & \nodata & 15.73 (0.07) & \nodata & \nodata & 8.53 (0.05) & \nodata & MG55 & CCNAA   \\
194 & 0.29 & 5:32:37.7 & -67:08:31.1 & 17.34 (0.12) & 16.66 (0.11) & \nodata & \nodata & \nodata & \nodata & MG58 & NNNNN   \\
195 & 0.48 & 5:32:56.0 & -65:16:42.0 & 14.73 (0.03) & 14.11 (0.03) & 13.27 (0.03) & 12.90 (0.03) & 9.31 (0.07) & \nodata & MG59 & AAAAA   \\
196 & 0.57 & 5:33:30.9 & -69:08:13.3 & 16.11 (0.07) & 15.51 (0.06) & \nodata & 12.82 (0.06) & 9.29 (0.10) & \nodata & MG60 & CCCCC   \\
197 & 0.99 & 5:33:13.1 & -72:36:46.3 & 17.34 (0.12) & 16.52 (0.10) & \nodata & 13.89 (0.07) & 9.04 (0.06) & \nodata & MG61 & NNNAC  \\
198 & 1.25 & 5:34:36.5 & -68:18:27.9 & 16.27 (0.09) & 16.24 (0.10) & \nodata & \nodata & \nodata & \nodata & MG62 & CCCCC   \\
199 & 0.66 & 5:34:10.2 & -71:43:14.1 & 15.61 (0.05) & 15.51 (0.06) & \nodata & \nodata & \nodata & \nodata & MG63 & CCNNN   \\
200 & 1.86 & 5:35:12.8 & -67:37:58.0 & 16.44 (0.09) & 15.95 (0.08) & \nodata & \nodata & 9.39 (0.11) & \nodata & MG64 & AACCC   \\
201 & 1.42 & 5:35:10.3 & -69:39:38.9 & 15.92 (0.07) & 14.74 (0.04) & \nodata & 12.53 (0.07) & 6.74 (0.03) & \nodata & MG65 & CCCCC \\
202 & 0.33 & 5:37:59.5 & -65:58:50.0 & 16.96 (0.10) & 16.56 (0.10) & 14.83 (0.08) & \nodata & \nodata & \nodata & MG66 & NAAAN  \\
203 & 0.63 & 5:37:38.0 & -71:41:37.9 & \nodata & 16.58 (0.11) & \nodata & \nodata & 9.14 (0.07) & \nodata & MG67 & NNNAA   \\
204 & 0.49 & 5:38:19.5 & -68:58:37.2 & 15.49 (0.05) & 15.12 (0.05) & \nodata & 11.77 (0.04) & 8.03 (0.05) & \nodata & MG68 & CCCCC   \\
205 & 1.42 & 5:39:15.5 & -66:49:43.5 & 18.00 (0.18) & 17.08 (0.14) & \nodata & \nodata & \nodata & \nodata & MG69 & NNNNN  \\
207 & 0.97 & 5:39:54.9 & -66:34:13.1 & 17.05 (0.11) & 16.33 (0.09) & \nodata & \nodata & 9.45 (0.14) & \nodata & MG71 & NAANA   \\
208 & 1.3 & 5:40:20.3 & -67:02:02.1 & \nodata & 16.47 (0.10) & \nodata & 14.35 (0.11) & 10.01 (0.16) & \nodata & MG72 & NNNAA   \\
211 & 1.97 & 5:42:15.4 & -68:48:55.7 & 17.16 (0.12) & \nodata & \nodata & \nodata & \nodata & \nodata & MG75 & NNNNN   \\
212 & 0.75 & 5:42:24.2 & -69:53:05.1 & 16.13 (0.07) & 15.58 (0.07) & 15.50 (0.17) & 13.27 (0.07) & 7.85 (0.04) & \nodata & MG76 & CCCAA \\
213 & 1.56 & 5:43:47.6 & -68:38:35.1 & 17.81 (0.18) & 16.30 (0.09) & \nodata & \nodata & 10.22 (0.20) & \nodata & MG77 & NNCCC   \\
214 & 0.29 & 5:46:25.4 & -67:50:02.8 & 17.07 (0.11) & 16.12 (0.08) & \nodata & \nodata & \nodata & \nodata & Mo40 & NACAN   \\
215 & 0.59 & 5:45:59.7 & -71:19:04.9 & 13.34 (0.02) & 13.39 (0.02) & 13.41 (0.03) & 13.41 (0.06) & \nodata & \nodata & MG79 & CCCCC \\
217 & 0.31 & 5:51:00.1 & -67:58:45.6 & 17.38 (0.12) & 16.35 (0.09) & \nodata & \nodata & \nodata & \nodata & MG81 & NNNEN   \\
218 & 0.73 & 5:53:11.9 & -71:28:57.0 & 16.46 (0.08) & 15.82 (0.08) & \nodata & \nodata & \nodata & \nodata & MG82 & AANAN   \\
224 & 1.69 & 4:34:46.7 & -69:23:19.6 & \nodata & 16.73 (0.11) & \nodata & \nodata & \nodata & \nodata & Mo2 & NNNAN   \\
225 & 1.17 & 4:36:14.4 & -70:31:03.0 & 18.20 (0.19) & \nodata & \nodata & \nodata & \nodata & \nodata & Mo3 & NNNNN   \\
227 & 0.88 & 4:45:19.9 & -67:34:33.4 & 17.44 (0.13) & 16.61 (0.11) & \nodata & 14.39 (0.09) & \nodata & \nodata & Mo5 & NNNEN \\
228 & 0.83 & 4:46:33.1 & -70:12:46.4 & 17.27 (0.12) & 16.51 (0.10) & 15.34 (0.12) & 13.53 (0.05) & 9.31 (0.08) & \nodata & Mo6 &  AAAAA \\
231 & 1.98 & 4:52:11.4 & -70:12:46.4 & 10.22 (0.00) & 9.44 (0.00) & 8.86 (0.00) & 8.03 (0.00) & 5.84 (0.03) & \nodata & Mo9 & AAAAA \\
232 & 0.39 & 4:55:45.4 & -65:48:34.7 & 17.10 (0.11) & 16.68 (0.11) & 14.90 (0.08) & 13.23 (0.04) & 8.74 (0.06) & \nodata & Mo10 & AAAAA  \\
233 & 0.61 & 4:58:03.2 & -70:25:08.8 & 16.83 (0.09) & 15.99 (0.07) & 14.72 (0.07) & 13.28 (0.04) & \nodata & \nodata & Mo11 & AAAAA \\
234 & 0.42 & 4:58:37.0 & -69:35:46.5 & \nodata & 15.42 (0.06) & \nodata & 13.18 (0.04) & \nodata & \nodata & Mo12 & CCCCN   \\
238 & 0.78 & 5:06:14.9 & -69:48:24.7 & \nodata & 14.93 (0.05) & 14.19 (0.05) & 13.16 (0.05) & \nodata & \nodata & Mo16 & CCCCA   \\
239 & 0.69 & 5:07:25.5 & -67:28:49.5 & 16.68 (0.09) & 15.95 (0.07) & 14.55 (0.06) & 13.33 (0.05) & \nodata & \nodata & Mo17 & CCAAN   \\
240 & 0.91 & 5:09:11.2 & -67:34:01.7 & 15.75 (0.06) & 15.12 (0.05) & \nodata & 12.60 (0.03) & 7.99 (0.04) & \nodata & Mo18 & AAAAA   \\
241 & 0.76 & 5:11:00.5 & -70:05:04.8 & 16.92 (0.11) & 16.33 (0.10) & \nodata & 13.63 (0.06) & 8.69 (0.06) & \nodata & Mo19 & CCCAA \\
242 & 0.89 & 5:11:10.8 & -71:10:26.4 & \nodata & 16.04 (0.08) & \nodata & \nodata & \nodata & \nodata & Mo20 & NNNAA   \\
243 & 0.21 & 5:19:04.1 & -64:44:38.1 & \nodata & 16.41 (0.09) & \nodata & \nodata & \nodata & \nodata & Mo21 & NAAAN   \\
245 & 1.45 & 5:21:52.3 & -69:43:18.0 & 16.16 (0.10) & \nodata & \nodata & \nodata & \nodata & \nodata & Mo23 & CCCCC   \\
246 & 0.33 & 5:22:53.2 & -71:05:40.7 & 16.85 (0.10) & 16.03 (0.08) & 14.78 (0.08) & 13.09 (0.04) & 9.22 (0.09) & \nodata & Mo24 & CAAAA   \\
247 & 0.67 & 5:26:02.6 & -72:31:02.5 & 17.00 (0.10) & 16.17 (0.08) & 14.79 (0.07) & 13.54 (0.05) & \nodata & \nodata & Mo25 & AAAAA   \\
248 & 1.73 & 5:28:01.5 & -70:13:30.1 & \nodata & \nodata & \nodata & 12.86 (0.08) & 8.78 (0.08) & \nodata & Mo26 & CCCCC \\
249 & 0.98 & 5:29:16.7 & -69:37:18.1 & 14.88 (0.04) & 14.75 (0.04) & 14.11 (0.06) & 12.80 (0.04) & 9.75 (0.19) & \nodata & Mo27 & CCCCC \\
250 & 0.96 & 5:29:18.4 & -70:23:49.7 & 14.46 (0.03) & 14.70 (0.04) & 13.74 (0.04) & 12.21 (0.03) & 8.10 (0.04) & \nodata & Mo28 & CCCCA  \\
252 & 0.84 & 5:31:35.3 & -69:23:46.4 & 17.11 (0.15) & \nodata & \nodata & \nodata & 9.84 (0.16) & \nodata & Mo30 & NNNNC   \\
254 & 0.39 & 5:32:05.2 & -69:57:27.4 & \nodata & 15.72 (0.07) & \nodata & \nodata & \nodata & \nodata & Mo32 & CCAAN   \\
255 & 0.98 & 5:32:09.3 & -70:24:41.5 & 13.74 (0.02) & \nodata & \nodata & \nodata & 9.29 (0.08) & \nodata & Mo33 & CCCCC   \\
258 & 1.26 & 5:38:53.6 & -69:57:55.7 & 14.93 (0.04) & 15.16 (0.06) & \nodata & 12.71 (0.10) & \nodata & \nodata & Mo36 & CCCCC   \\
259 & 0.51 & 5:39:14.5 & -70:00:18.6 & 15.84 (0.07) & 15.49 (0.06) & 13.36 (0.04) & \nodata & \nodata & \nodata & Mo37 & CCCCC   \\
260 & 0.82 & 5:40:32.3 & -68:44:47.6 & 16.51 (0.08) & 15.69 (0.07) & \nodata & 12.89 (0.05) & 9.80 (0.15) & \nodata & Mo38 & CCCCC   \\
261 & 0.63 & 5:42:41.1 & -70:05:49.1 & 16.78 (0.10) & 15.72 (0.07) & \nodata & \nodata & \nodata & \nodata & Mo39 & NACCC   \\
262 & 0.71 & 5:46:25.3 & -71:23:22.3 & 17.12 (0.11) & 16.19 (0.09) & \nodata & 14.15 (0.09) & \nodata & \nodata & Mo40 & CCCAA   \\
264 & 0.54 & 5:55:14.6 & -66:50:24.6 & 16.67 (0.09) & 15.76 (0.07) & \nodata & 13.69 (0.06) & 9.44 (0.09) & \nodata & Mo42 & AACAA   \\
266 & 1.32 & 6:01:43.3 & -68:00:35.0 & \nodata & 16.12 (0.08) & \nodata & \nodata & \nodata & \nodata & Mo44 & CCAAN   \\
277 & 0.32 & 5:07:30.8 & -69:08:07.0 & \nodata & \nodata & \nodata & 9.29 (0.01) & 4.18 (0.02) & \nodata & SMP26 & CCCCA \\
\enddata
\tablenotetext{a}{Distance in arcsec between the position determined from
the 3.6 $\mu$m photometry and the \citet{leisy97} catalog position.}
\tablenotetext{b}{Characteristics of the source and field near the nebula.  
A=well defined, isolated point source, B=blended with other nearby point
source, C=complex background or distribution of many nearby point sources,
E=extended source, N=no source visible or too faint to determine whether 
extended or pointlike.}
\tablenotetext{c}{The source identifications given by \citet{leisy97} for
the objects in their catalog.  Abbreviations are J: Jacoby \citet{jacoby80},
MG: \citet{morgan92},
Mo: \citet{morgan94}, Sa: \citet{sand84}, SMP: \citet{smp78}.
}
\tablenotetext{d}{Because it is extended in the IRAC images, the
photometry for LMC 92 (SMP 93) was performed by summing the flux above the
background in a 25\farcs62 $\times$ 17\farcs08 box centered on the core.}
\end{deluxetable}
\clearpage
\end{landscape}

\end{document}